\documentclass[12pt,a4paper]{article}
\usepackage{a4wide,hyperref}
\usepackage{epsfig}
\usepackage{latexsym}
\usepackage{amsfonts}
\usepackage{amsmath}
\usepackage{amsthm}
\usepackage{amssymb}
\usepackage{amsbsy}
\usepackage{multirow}
\usepackage{slashed}
\usepackage{color}
\usepackage{mathrsfs}
\usepackage{subfigure}
\usepackage[font={footnotesize,sl},bf]{caption}
\usepackage{wrapfig}
\usepackage{enumerate}

\usepackage{cite}

\def\caln         {{\cal N}}

\newsavebox{\uuunit}
\sbox{\uuunit}
    {\setlength{\unitlength}{0.825em}
     \begin{picture}(0.6,0.7)
        \thinlines
        \put(0,0){\line(1,0){0.5}}
        \put(0.15,0){\line(0,1){0.7}}
        \put(0.35,0){\line(0,1){0.8}}
       \multiput(0.3,0.8)(-0.04,-0.02){12}{\rule{0.5pt}{0.5pt}}
     \end {picture}}
\newcommand {\unity}{\mathord{\!\usebox{\uuunit}}}

\newcommand{\beq}{\begin{eqnarray}}
\newcommand{\eeq}{\end{eqnarray}}

\newcommand{\eal}[1]{\begin{equation} \begin{aligned} #1 \end{aligned}\end{equation}}

\newcommand{\be}{\begin{equation}}
\newcommand{\ee}{\end{equation}}
\newcommand{\bea}{\begin{eqnarray}}
\newcommand{\eea}{\end{eqnarray}}
\newcommand{\bean}{\begin{eqnarray*}}
\newcommand{\eean}{\end{eqnarray*}}

\def\rmp{{\Gamma}}
\def\rmq{{\Gamma}}
\def\rmh{{h}}

\def\to{\rightarrow}

\def\sF{{{ F}\!\!\!\!\hskip.8pt\hbox{\raise1pt\hbox{/}}\,}}
\def\som{{{ \omega}\!\!\!\!\hskip.8pt\hbox{\raise1pt\hbox{/}}\,}}
\def\sJ{{{\rm J}\!\!\!\!\hskip.8pt\hbox{\raise1pt\hbox{/}}\,}}

\newcommand{\bdm}{\begin{displaymath}}
\newcommand{\edm}{\end{displaymath}}

\begin{document}
\setcounter{tocdepth}{2}

%
%
%
%
%

\begin{titlepage}

\numberwithin{equation}{section}
\begin{flushright}
\small
IPhT-T15/217\\
MCTP-15-32

\normalsize
\end{flushright}
\vspace{0.8 cm}

\begin{center}

\hspace*{-1cm}\mbox{\LARGE \textbf{Tunneling into Microstate Geometries:}}\\ \vspace{0.5cm}
{\Large \it Quantum Effects Stop Gravitational Collapse}

\medskip

\vspace{1.2 cm} {\large Iosif Bena$^a$, Daniel R.\ Mayerson$^{b,c}$, Andrea Puhm$^d$, Bert Vercnocke$^b$}\\

\vspace{1cm} {$^a$Institut de Physique Th\'eorique, Universit\'e Paris Saclay,\\ CEA, CNRS, F-91191 Gif-sur-Yvette, France}

\vspace{0.5cm} {$^b$Institute for Theoretical Physics, University of Amsterdam,\\
 Science Park 904, Postbus 94485, 1090 GL Amsterdam, The Netherlands}
 
 \vspace{0.5cm} {$^c$Department of Physics and Michigan Center for Theoretical Physics,\\ University of Michigan,
 450 Church Street, Ann Arbor, MI 48109-1020, USA}

\vspace{0.5cm} {$^d$ Department of Physics, University of California, Santa Barbara, CA 93106, USA}

\vspace{0.5cm}

\vspace{.5cm}
iosif.bena @ cea.fr, drmayer @ umich.edu, \\
puhma @ physics.ucsb.edu,
bert.vercnocke @ uva.nl\\

\vspace{2cm}

\textbf{Abstract}
\end{center}


Collapsing shells form horizons, and when the curvature is small classical general relativity is believed to describe this process arbitrarily well. On the other hand, quantum information theory based (fuzzball/firewall) arguments suggest the existence of some structure at the black hole horizon. This structure can only form if classical general relativity stops being the correct description of the collapsing shell before it reaches the horizon size.
We present strong evidence that classical general relativity can indeed break down prematurely, by explicitly computing the quantum tunneling amplitude of a collapsing shell of branes into smooth horizonless microstate geometries. We show that the amplitude for tunneling into microstate geometries with a large number of topologically non-trivial cycles is parametrically larger than $e^{-S_{BH}}$, which indicates that the shell can tunnel into a horizonless configuration long before the horizon has any chance to form. We also use this technology to investigate the tunneling of M2 branes into LLM bubbling geometries.

\end{titlepage}

\newpage

\tableofcontents

\newpage

\section{Introduction and Summary}\label{sec:intro}

The argument that black holes must have a certain nontrivial structure at the horizon to avoid violation of quantum mechanics \cite{Mathur:2009hf,Almheiri:2012rt} (see also \cite{Braunstein:2009my}) is surprising on several fronts. The first is that, wherever this structure comes from, it must have very peculiar properties: since the horizon is a null surface, this structure cannot come from ordinary particles, which would travel at the speed of light and have infinite mass; nor can it come from massless particles, which dilute in a few horizon crossing times and must be constantly replenished. The second is that this structure must be able to account for the Bekenstein-Hawking entropy of the black hole, and thus must have nontrivial degrees of freedom to give rise to this entropy. The third, and perhaps most surprising one, is that an infalling shell of dust which initially undergoes gravitational collapse \`{a} la Oppenheimer-Snyder~\cite{Oppenheimer:1939ue} must somehow turn into this structure at the moment it crosses the 
Schwarzschild radius, regardless of how low its curvature is and how adept one might hope classical 
general relativity to be for describing its physics.

The first two points are addressed by the black hole microstate or fuzzball solutions that one constructs in string theory (see \cite{Mathur:2005zp,Bena:2007kg,Balasubramanian:2008da,Skenderis:2008qn,Mathur:2008nj,Chowdhury:2010ct,Bena:2013dka} for reviews). The purpose of this paper is to address the last question and show how infalling matter can tunnel into smooth geometric microstates that give rise to structure at the scale of the would-be horizon. 

\subsection{Motivation}

In string theory, and in string theory only, one can build structure at the scale of the horizon of black holes with a large horizon area \cite{Bena:2005va,Berglund:2005vb,Bena:2006kb,Bena:2010gg,Bena:2015bea}  by solving the equations of motion of the low-energy limit of a UV-complete theory. This structure has the desired features: it is kept from collapsing by fluxes wrapping topologically-nontrivial cycles, and for extremal supersymmetric black holes there is evidence that it has enough degrees of freedom to reproduce the growth with charges of the Bekenstein-Hawking entropy \cite{Bena:2014qxa}. 

Note that if one tries to construct such structure-at-the-horizon in other gravity theories, one is almost automatically guaranteed to fail. For example, in four-dimensional gravitational theories there are ``no solitons without horizon'' theorems that guarantee that no smooth horizonless solutions with black hole charges can be built \cite{Nogo1,Nogo2,Nogo3,Breitenlohner:1987dg,Nogo4}. 
Making the solutions singular and adding extra matter to source the singularities does not work 
either, as the matter is generically not stiff enough to prevent its collapse and the subsequent formation of a black hole. It is only by going to higher dimensions and exploiting the extra fluxes that are required by string theory in order to have a consistent quantum theory of gravity, that one can create such solitons \cite{Gibbons:2013tqa,Kunduri:2013vka,Haas:2014spa,deLange:2015gca}.

However, despite the existence of such structures and their ability to carry the black hole entropy, the question still remains of how to convince a very large collapsing shell of dust, whose Schwarzschild radius can be of the size of the Galaxy and whose curvature is much smaller than that on the surface of the Earth, to transform itself into this structure around the moment it reaches its Schwarzschild radius. Of course, as the size of the shell becomes smaller and smaller the curvature grows and, near the Planck scale, one expects quantum gravity effects to take over and possibly transform the shell into a string theory-structure of the kind discussed in  \cite{Mathur:2005zp,Bena:2007kg,Balasubramanian:2008da,Skenderis:2008qn,Mathur:2008nj,Chowdhury:2010ct,Bena:2013dka}. But this will be too late: the horizon of non-extremal black holes is in the causal past of the moment when the curvature of the shell is large enough for general relativity to break down. Hence, unless one goes backwards in time and 
destroys several million of years of past history, one cannot create any structure 
at the horizon.

Only one proposal has so far been put forth to explain how a collapsing shell of dust might transform itself when reaching the size of the would-be event horizon into a horizonless configuration or a fuzzball. In \cite{Mathur:2008kg}  (see also \cite{Mathur:2009zs}) Mathur considered a shell of matter of mass $M$, collapsing into its (classical) Schwarzschild black hole horizon. This shell can quantum tunnel into a given horizonless configuration with an estimated tunneling amplitude
\be \label{eq:introGamma} \Gamma \sim e^{-S_{\rm tunnel}}\quad \text{with} \quad S_{\rm tunnel} = \alpha S_{BH}\,, \ee
where dimensional analysis was used to obtain $S_{\rm tunnel} \sim \int \sqrt{-g}R \sim M^2 \sim S_{BH}$. The assumption that the curvature scale of the fuzzball is given by the black hole length scale sets the proportionality constant $\alpha$ to be of order one.
Although the tunneling rate~\eqref{eq:introGamma} is a very small number, if one assumes that the so-called fuzzball proposal is correct and the entropy of the black hole comes entirely from horizonless configurations, the total number of states the shell can tunnel into is very large:
\be 
\label{eq:introN} \mathcal{N} = e^{S_{BH}} 
\,.
\ee
The two exponentials in (\ref{eq:introGamma}) and (\ref{eq:introN}) play off against each other, and if $\alpha\leq 1$ tunneling into fuzzball states is \emph{fast} and takes place before a horizon can form. In a subsequent paper, \cite{Kraus:2015zda} Kraus and Mathur argued that $\alpha$ should be equal to $1$, by estimating that the probability for a shell to tunnel into a fuzzball is the same same as that for a shell of dust to be emitted by a Schwarzschild black hole, which is exactly the exponential of the negative of the entropy difference $\Gamma \sim \exp(-\Delta S_{BH})$. By extrapolating this result one may then argue that the probability of the the collapsing shell to tunnel into a particular entropy-less microstate is $\Gamma \sim \exp(- S_{BH})$ and therefore a collapsing shell of dust would tunnel into some fuzzball with probability one, rather than forming a horizon.

In this paper we are able to put calculational flesh on the proposal of  \cite{Mathur:2008kg}, by directly computing the amplitude~\eqref{eq:introGamma} for a collapsing shell of branes to tunnel into some of the black hole microstate solutions that have been explicitly constructed in the past~\cite{Bena:2005va,Bena:2006kb,Berglund:2005vb}. Our results confirm some of the expectations of the proposal of \cite{Mathur:2008kg} and also show that some of the assumptions in the analysis of \cite{Kraus:2015zda} are not necessary for the validity of the overall argument.  We will discuss in detail in Section \ref{Kraus-Mathur} the differences between our method to compute tunneling of collapsing shells and the method used in \cite{Kraus:2015zda}.

\subsection{Summary of the results}

The microstate geometries whose tunneling we analyze are solutions with nontrival topology, and charges dissolved in flux. However, the technology we develop to study brane-flux tunneling is much more general, and can be easily adapted to other solutions where the branes can tunnel into fluxes. This technology has been used in the past to study the tunneling of metastable probes with anti-D3 charge~\cite{Kachru:2002gs} in the KS solution~\cite{Klebanov:2000hb} and for the tunneling of probe anti-M2 branes~\cite{Klebanov:2010qs} in the CGLP solution \cite{Cvetic:2000db}. Here we will apply this technology for general solutions of  string theory and M-theory where branes are transitioned into flux and topology. 

The rate for the tunneling decay per unit volume is given by \cite{Coleman:1977py}:\footnote{We will be interested in the leading-order term in~\eqref{eq:Gamma} and henceforth set $\hbar=1$.}
\begin{equation}\label{eq:Gamma}
  \Gamma/V = A e^{-B/\hbar}\left(1+\mathcal{O}(\hbar)\right)\,.
\end{equation}
The tunneling parameter $B$, which can be determined from the Euclidean action, is the focus of our paper.
We compute the exponent $B$ for tunneling processes that create non-trivial topology, by placing probe strings or branes in multi-center backgrounds. We consider probes \cite{Bena:2011fc, Bena:2012zi} in microstate geometries of three-charge black holes~\cite{Bena:2005va,Bena:2006kb,Berglund:2005vb} and also to study the tunneling of the recently found probe metastable states~\cite{Massai:2014wba} in the Lin-Lunin-Maldacena (LLM) geometries~\cite{Lin:2004nb} dual to the mass-deformed M2 brane theory \cite{Bena:2000zb}.
 In their M-theory descriptions, these two classes of geometries share several similarities. For instance, they both account for the entropy of the dual theory: either the Bekenstein-Hawking entropy of the black hole or the number of vacua of the M2 brane theory \cite{Gomis:2008vc,Cheon:2011gv}. However, they differ in one key feature which leads to a very different tunneling behavior. For black hole microstates, the tunneling calculation reduces to a quantum-particle tunneling problem which can be tackled analytically. For the LLM geometries, one has to consider an $O(d)$ invariant Euclidean tunneling action which can only be treated analytically after an approximation.

\subsubsection{Black hole microstate geometries}

Starting from a collapsing shell of branes, we imagine forming smooth multi-center geometries in a stepwise quantum tunneling process, where we form a new center, and hence a new topologically-nontrivial cycle, at every step by tunneling an amount of the initial branes into fluxes. At each step, we treat the tunneling branes as probes, which allows us to calculate their tunneling potential from the on-shell Euclidean (probe) action explicitly. Using this stepwise process, we determine the number of topologically non-trivial cycles at the end of the tunneling process. Note that since we are treating the branes as probes we can only tunnel a small amount of them at every step. However, we can still end up with a large number of bubbles by successively tunneling small amounts of branes.

We find that the tunneling amplitude to a final state with $N$ centers scale as:
\be \label{eq:introGammascaleN}
\Gamma \sim \exp(-\alpha_0 N^{-\beta}S_{BH})\,,
\ee
where $\alpha_0$ is a microstate-dependent number and the exponent $\beta$ is positive for all the solutions we have considered\footnote{For the non-scaling and scaling solutions we considered we found, respectively, $\beta=3/2$ and $\beta=0.93$.}.
and the black hole entropy is $S_{BH}=2\pi\sqrt{Q^3}$, with $Q$ the electric charge of the black hole.

The key information lies in the prefactor $\alpha \equiv \alpha_0 N^{-\beta}$, from which we extract a universal feature of the tunneling amplitude into multi-center solutions: its dependence on the number of centers, $N$.
Since the power of $N$ appearing in the exponential is negative ($\beta>0$), the result~\eqref{eq:introGammascaleN} implies that the tunneling amplitude is enhanced when the number of centers is large. It is also important to emphasize that, even if we do our tunneling calculation for bubbling solutions that have $U(1) \times U(1)$ isometry in five dimensions, the tunneling rate will be the same for the black hole microstate solutions one constructs by wiggling these solutions to form superstrata and other wiggly objects; this happens because there is no potential barrier between various wiggly solutions. 

This being said, one should also emphasize that we do not know yet whether the microstate geometries that give rise to the full black hole entropy have very few bubbles of very many.
From the study of superstrata \cite{Bena:2013dka,Bena:2011uw} it may appear that the solutions that can reproduce the growth with charges of the black hole entropy should have only a few centers, although in \cite{Bena:2015bea} it was argued that the typical superstrata may look more like multi-center solutions than double-centered ones. Similarly, from the study of quiver quantum mechanics one can reproduce \cite{Denef:2007vg} the charge growth of the supersymmetric four-dimensional four-charge black hole entropy from the pure-Higgs states \cite{Bena:2012hf} of three-center configurations; however, this entropy is a very small fraction of the black hole entropy\footnote{For the three-center scaling solution of \cite{Bena:2007qc}, it represents $4\%$ of the black hole entropy \cite{Martinec:2015pfa}.} and it is not clear whether solutions with more centers could carry larger fractions of the black hole entropy.

Hence, the detailed aspects of the physics of a collapsing shell would be determined by the interplay between the increased tunneling rates into multi-bubble solutions, and the possible larger number of tunneling end-points available in few-bubble solutions. Nevertheless, from our calculations it is clear that the tunneling rate both into solutions with small and large numbers of bubbles will be more than enough to ensure that the shell tunnels before a horizon can form,  irrespective of the details of which solutions carry more entropy.

The amplitude~\eqref{eq:introGammascaleN} describes the direct tunneling of a shell of branes into one individual microstate. However, to compute the total tunneling amplitude of the shell into this state one also has to sum over two-step tunneling processes (in which the shell tunnels to a different microstate and then tunnels from that microstate to the one we consider) as well as three-step tunneling processes, etc. Furthermore, since there are $\mathcal{N} \equiv e^{S_{BH}}$ possible intermediate steps, this two-step tunneling probability could be larger than the one-step one, thus invalidating our semiclassical approximation\footnote{We would like to thank Juan Maldacena for bringing this point to our attention.}. To argue that this does not happen we can estimate the multi-step contributions to the tunneling process. The one-step probability is given by~\eqref{eq:Gamma}. The tunneling probability for a particular two-step process is $\Gamma^2$ and because of the existence of $\mathcal{N}$ possible 
intermediate steps his should be naively multiplied by a factor of $\mathcal{N}$. However, the amplitudes coming from the intermediate steps have different phases, and the resulting amplitude will be reduced by a factor of $\sqrt{\mathcal{N}}$ because of destructive interference\footnote{Remember that after a random walk of $\mathcal{N}$ steps the average displacement is of order $\sqrt{\mathcal{N}}$}. Hence, the two-step tunneling amplitude is of order  $\Gamma^2 \sqrt{\mathcal{N}} $. Similarly, the three-step amplitude is of order $\Gamma^3$ for a particular tunneling path, but because there are $\mathcal{N\, }^2$ possible possible paths with quantum interference this will be multiplied by a factor of $\sqrt{\mathcal{N\,}^2} $. It is not hard to see therefore that the full tunneling amplitude will be given by a convergent sum  $\Gamma(1 + \Gamma \sqrt{\mathcal{N}} + \Gamma^2 \mathcal{N} + ...)$ and thus the semiclassical approximation is valid.

\subsubsection{LLM geometries}
The action that describes the tunneling of branes into flux and topology has been computed in the literature before \cite{Kachru:2002gs,Klebanov:2010qs} to estimate by now well-known decay rates of metastable states from a Brown-Teitelboim process \cite{Brown:1988kg}. Along similar lines, we estimate the lifetime of metastable brane configurations~\cite{Massai:2014wba} in the LLM geometries in string and M-theory. 

The key difference to the black hole microstate geometries is that the probes that tunnel in the LLM geometries extend along one or two non-compact spatial directions (as opposed to the branes in the black hole microstates, which only wrap compact directions). This implies that we will need to compute the $O(3)$ or $O(2)$ Euclidean actions~\cite{Coleman:1977py} corresponding to, respectively, tunneling of probe M2 branes in the M-theory geometry or of probe F1 strings in the reduced IIA geometry.
We find that the amplitude for metastable LLM configurations to tunnel into their ground state scales as:
\be \Gamma_{M2} \sim \exp(-\alpha_{M} \,\mu^6/q^2) \quad \text{(11D)}\,, \qquad \Gamma_{F1} \sim \exp(-\alpha_{IIA} \, \mu^4/|q|) \quad \text{(10D)}\,,\ee
where $\mu$ is the mass deformation parameter that sets the scale of the four-cycles in the bubbling LLM solutions and $q$ is the charge of the tunneling probe M2 branes/F1 strings. The parameters $\alpha_{M}$ and $\alpha_{IIA}$ will be computed in section \ref{sec:LLM}.  The main difference with the black hole microstate tunneling events is that the exponent $B$ does not scale linearly with the charges, but is inversely proportional to $q^{d}$, with $d$ the number of non-compact directions \cite{Brown:1988kg,Coleman:1977py}. The dependence on $q$ and $\mu$ differs between the 10D and 11D description since the $O(2)$ bounce computes tunneling per unit length while the $O(3)$ bounce computes tunneling per unit area. For both descriptions, quantum tunneling is suppressed for small charge $|q|$ and large mass deformation $\mu$.  

\subsection{Organization of this paper}
In section \ref{sec:basics} we review the basics of tunneling rates for particles and the $O(d)$ generalization to strings and branes; readers familiar with these (standard) methods can skip this section. In section \ref{sec:multicenter} we consider the tunneling of branes into multi-center microstate geometries of the three-charge maximally-spinning black hole, and investigate the scaling of the tunneling amplitude with the number of centers $N$. In section \ref{sec:LLM}, we consider the $O(2)$ and $O(3)$ tunneling rates of metastable states in LLM-type backgrounds in string theory and M-theory. Details of the computations of section \ref{sec:multicenter} and \ref{sec:LLM} can be found in appendices \ref{app:SUGRAconventions}, \ref{app:multicenter}, and \ref{app:LLM}.

\section{Quantum Tunneling of Particles and Branes}\label{sec:basics}
In this section, we review the tunneling decay rate for a quantum particle tunneling through a potential barrier as detailed in Coleman's seminal paper \cite{Coleman:1977py}, and the generalization to strings and branes. First, we discuss a
charged particle in a curved background with a position-dependent mass, which will be relevant for section \ref{sec:multicenter}. Then we review the generalization to an $O(d)$ symmetric Euclidean action that describes the nucleation of a bubble of true vacuum mediating the decay of a metastable vacuum, relevant for section~\ref{sec:LLM}. Finally, we make a comparison and emphasize the fundamental difference between the particle and $O(d)$-symmetric  strings and branes.

\subsection{Tunneling of particles}\label{ssec:general1D}

Calculating the tunneling rate of a quantum-mechanical particle in a $D$-dimensional target space is a standard  problem  discussed in \cite{Banks:1973ps}. In the language of \cite{Coleman:1977py}, the tunneling parameter $B$ can be determined from the on-shell Euclidean action:
\be B = S_E= \int_{t_i}^{t_f} dt\, L_E(x^k(t),\dot{x}^k(t))\,, \label{eq:S1D}
\ee
where $\dot{x}\equiv dx/dt$ and $k=0,\dots,D-1$. The integration is over the trajectory of the path that is a solution to the Euclidean equations of motion starting from an initial configuration at $t_i$ and ending in a final configuration at $t_f$ (a path of `least resistance'). Along this path, the Euclidean Hamiltonian is a constant of motion that can be chosen to be zero. From the general expressions for the Hamiltonian and the momentum conjugate to $x^k$:
\be  H_E = p_k\dot{x}^k - L_E=0\,,\qquad p_k = \frac{\partial L_E}{\partial \dot{x}^k}\,,\ee
we get $L_E = p_k\dot{x}^k$.

For a non-relativistic particle in a potential, that is described by the Euclidean Lagrangian $L_E = (m/2) g_{k\ell}\dot x^k \dot x^\ell  + V(x)$, one finds that $p_k\dot{x}^k = |p| |\dot x|$. The above integral \eqref{eq:S1D} becomes
\begin{equation} \label{eq:finalB1D} 
 B =  \int_{\vec x_i}^{\vec x_f} |dx|\, |p (x)|\,,
\end{equation}
where the integral is taken over a path of least resistance and $|p| \equiv \sqrt{g^{k\ell}p_k p_\ell}, |dx| \equiv \sqrt{g_{k\ell} dx^k dx^\ell}$. In other words, instead of finding the entire trajectory by solving the equations of motion, we can simply integrate the norm of the Euclidean momentum $|p|$ in position space from the starting point $\vec x_i$ to the endpoint $\vec x_f$ of the tunneling process. Note that $H_E = 0$ implies further that $|p|=\sqrt{2mV}$ and therefore we recover the result of the usual WKB approximation \cite{Coleman:1977py}.

\begin{figure}[ht!]
\begin{center}
 \includegraphics[width=0.4\textwidth]{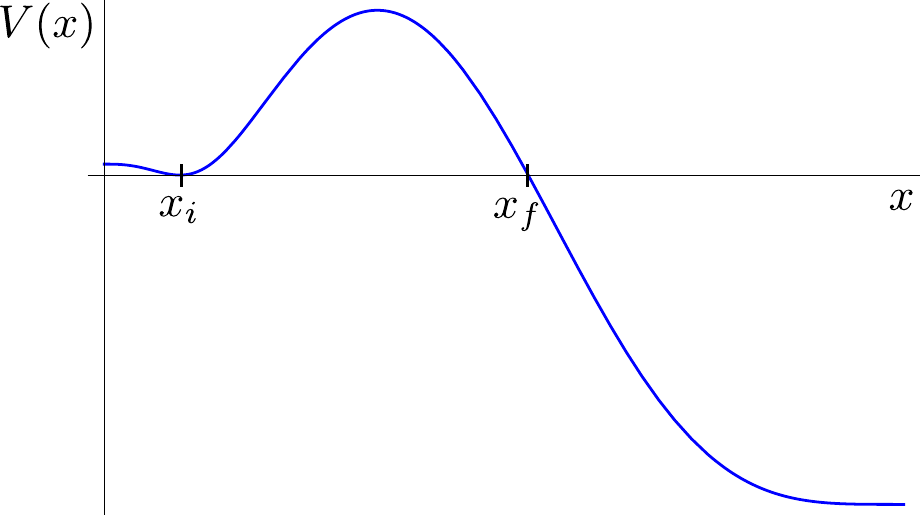}
 \end{center}
 \caption{We calculate the tunneling rate for a particle tunneling through a potential barrier. After tunneling from $x_i$ to $x_f$ the particle experiences classical decay.
 }
 \label{fig:tunneling+particle}
\end{figure}

\paragraph{Relativistic particle.}
Consider now a (relativistic) particle with position-dependent mass $m(x)$ that  couples to a gauge field $A$ with charge $q$, in a curved $D$-dimensional background. We assume that the metric and all fields have a timelike symmetry $\partial_t$ and that the metric is static. The Euclidean action is:
\be S_E = \int m(x) ds  +  q\int  A(x)\,,
\ee
where integration is over the trajectory of the particle (pulling back $A$ appropriately). 
Taking the time-like coordinate $t$ as the affine parameter along the path we get:
\be
S_E = \int_{t_i}^{t_f} dt \,L_E\,, \qquad  L_E =  \left(g_{tt} + g_{ij} \dot x^i  \dot x ^j\right)^{1/2}  m(x) + q A_i(x)\dot{x}^i + q A_t(x)\ \,,
\ee
We will exclusively  consider paths along which $A_i \dot x^i =0$. Then $p_i$ is proportional to $g_{ij} \dot x^j$ and
we again find that the tunneling amplitude is given by integrating the norm of the momentum in position space:
\begin{equation}
B =  \int_{\vec x_i}^{\vec x_f} |dx|\, |p (x)|\,,
\end{equation}
and the norm of the momentum can be found from solving $H_E=0$:
\begin{equation}
\label{eq:pE1D} 
|p(x)| =|g_{tt}(x)|^{-1/2}\sqrt{|g_{tt}(x)| m(x)^2-(qA_t(x))^2} \,.
\end{equation}
As we explain below, we can use this expression to calculate the tunneling amplitudes in black hole microstate geometries in section~\ref{sec:multicenter}.

\subsection{Bubble nucleation for extended objects with \texorpdfstring{$O(d)$}{} symmetry}\label{ssec:generaldD}

To describe the tunneling of extended objects like strings and branes wrapping non-compact cycles we have to generalize the Euclidean action for a particle of section~\ref{ssec:general1D}. We have to promote the trajectory $x(t)$ to a function of the worldvolume coordinates $x(\sigma_i)$ with $i=0,...,d-1$.

We consider the $O(d)$ invariant tunneling process\footnote{The $O(d)$-symmetric tunneling process is favored over a non-symmetric one, as mentioned in \cite{Coleman:1977py} and proven in \cite{Coleman:1977th}.} where the trajectory only depends on the Euclidean radius $R=\sqrt{\sum_i (\sigma_i)^2}$. Then, the tunneling parameter $B$ can be obtained from the Euclidean action:
\begin{equation}
 B=S_E=\int_{S^{d-1}} d\Omega_{d-1} \int_{R_i}^{R_f} dR\, R^{d-1} L_E(x(R),\partial_R x(R))\,.
\end{equation}
The (radial) Hamiltonian and momentum conjugate to $x$ are:
\begin{equation}
 H_E=p \partial_R x - L_E\,, \qquad p = \frac{\partial L_E}{\partial(\partial_R x)}\,.
\end{equation}

As for the particle, the Hamiltonian is a constant of motion that we can choose $H_E=0$, so that $L_E=p \partial_R x$. The expression for the tunneling parameter becomes
\begin{equation}\label{eq:finalBdD}
 B=\int_{S^{d-1}} d\Omega_{d-1} \int_{R_i}^{R_f} dR\, R^{d-1} \,\partial_R x \,p(x)\,.
\end{equation}
An important difference between this $O(d)$ invariant action and the action~\eqref{eq:finalB1D} is that now we can no longer  replace the integration over the Euclidean radius $R$ (which plays the role of $t$ in (\ref{eq:finalB1D})) by an integration over the path $|dx|$ as we did in (\ref{eq:finalB1D}), because of the explicit factor of  $R^{d-1}$ in~\eqref{eq:finalBdD}. We will come back to the fundamental difference between tunneling of particles and extended objects in section \ref{ssec:1DvsdD}.

\paragraph{Relativistic effective string.}

We now generalize the results of relativistic particles to a relativistic effective string, so that we are considering $O(2)$-symmetric tunneling. The effective string can for instance descend from a D-brane (or M-brane) wrapping compact cycles and extending in one non-compact spatial direction.

The Euclidean action is:
\begin{equation}
 S_E=\int d^d\sigma (V_{DBI}+V_{WZ}), \qquad V_{DBI} = \sqrt{\det(g+A^{(1)})}, \qquad V_{WZ} = A^{(2)},
\end{equation}
where $V_{DBI}(x)$ and $V_{WZ}(x)$ correspond to the Dirac-Born-Infeld and Wess-Zumino Lagrangian. In principle, $g$ can be any symmetric two-tensor and $A^{(i)}$ can be any anti-symmetric two tensors; $g,A^{(i)}$ are all pulled back onto the worldvolume. The integration is over the trajectory $x(R)$ that connects the initial vacuum $x_i$ at large $R$ to the final vacuum $x_f$ at small $R$ via an instantonic domain wall at $R=R_*$. 

We have two worldvolume directions $\sigma^0,\sigma^1$ (so $R=\sqrt{(\sigma^0)^2+(\sigma^1)^2}$), and we take an embedding in spacetime as $t=\sigma^0,w=\sigma^1,x=x(R)$.  We further assume that the spacetime metric is diagonal in $t,x,w$ and moreover that $|g_{tt}|=g_{xx}$. There is one relevant component of each $A^{(i)}$, namely $A^{(i)}_{tw}$. All of these assumptions will be satisfied in the $O(2)$ tunneling event discussed later in section \ref{sec:LLM}. The $O(2)$ symmetric action is given by:
\begin{equation}\label{eq:Sgeneral}
 S_E = 2\pi  \int_{R_i}^{R_f} dR\, R\, L_E\,, \quad L_E=\left(1+\left|\frac{g_{xx}}{g_{tt}+(A^{(1)}_{tw})^2/g_{tt}}\right| \dot{x}^2\right)^{1/2} \sqrt{g_{tt}^2+(A^{(1)}_{tw})^2} + A^{(2)}_{tw} \,,
\end{equation}
where the dots  now stand for radial derivatives $\dot{x}\equiv \partial_R x$.

Solving $H_E=p \partial_R x - L_E=0$ we obtain the momentum conjugate to $x$:
 \begin{equation}\label{eq:pEgeneral}
  p(x)=\sqrt{\left|\frac{g_{xx}(x)}{g_{tt}(x)+A^{(1)}_{tw}(x)^2/g_{tt}(x)}\right|} \sqrt{g_{tt}^2+(A^{(1)}_{tw})^2-(A^{(2)}_{tw})^2} \,.
 \end{equation}

However, the tunneling action (\ref{eq:Sgeneral}) cannot be written as an integral in $x$-space of $p$, as we did for the tunneling of a particle, because of the presence of an extra factor of $R$.
We will use the expression (\ref{eq:Sgeneral}) in section~\ref{sec:LLM} for calculating the tunneling amplitude in LLM geometries.

\paragraph{Decay of metastable vacua and domain walls.}
Typically, one needs to resort to numerics in order to calculate the trajectory $x(R)$ and then integrate $p(x(R))$ given in (\ref{eq:pEgeneral}) to obtain $B$. Sometimes one can use an analytical approximation instead. When we consider the tunneling of a metastable vacuum into a stable vacuum, where the metastable vacuum only has a small excess energy compared to the stable one, we can use the thin domain wall approximation to evaluate
the action~\eqref{eq:Sgeneral}. In this approximation, for large $R$ the trajectory is approximated by the metastable vacuum $x_i$; for small $R$, it is approximately the true vacuum $x_f$; at the domain wall at $R_*$, the trajectory is approximately constant (since the domain wall is thin). We can consider the contribution to the Euclidean action of each region separately:
\begin{itemize}
\item $R\gg R_*$: We can estimate the energy at the metastable minimum from the effective potential:
\begin{equation}\label{eq:dDVeff}
 V_{\rm eff}(x)\equiv H_E(x,\partial_R x=0)=V_{DBI}+V_{WZ}\,.
\end{equation}
 Evaluating the effective potential at the minimum $x_i$ gives the following contribution to the action:
\begin{equation}
S_E\Big|_{R\gg R_*} = V_{S^{d-1}} \int_{R\gg R_*} dR R^{d-1} L_E\Big|_{\partial_R x=0}= -\frac{V_{S^{d-1}}}{d} R_*^d V_{\rm eff}(x_i) \,.
\end{equation}

 \item  $R\ll R_*$: Here, we have (approximately) the true vacuum, and since $V_{\rm eff}(x_f)=0$, the contribution to the Euclidean action is zero.
  
\item $R\approx R_*$: We take $R$ to be approximately constant so that we can take it outside of the integral and convert the latter into an integral over $x$:
  \begin{equation}
S_E\Big|_{R\approx R_*} = V_{S^{d-1}}\int_{R\approx R_*} dR R^{d-1} \,\partial_R x \,p(x)\approx  V_{S^{d-1}}R_*^{d-1} \int_{x_f}^{x_i} dx \,p(x) \equiv V_{S^{d-1}}R_*^{d-1} T_{\rm wall}\,,
\end{equation}
where in the last step we have defined the tension $T_{\rm wall}$ of the domain wall. In general, this tension is given by the action of a brane wrapping the contractible $S^{d-1}$ at $R=R_*$ (and possibly wrapping other compact directions).
\end{itemize}
We see that we are left with the action:
\be S_E(R_*) =   V_{S^{d-1}}R_*^{d-1} T_{\rm wall} -\frac{V_{S^{d-1}}}{d} R_*^d V_{\rm eff}(x_i), \ee
which we must still extremize with respect to $R_*$. The final result is:
\begin{equation}\label{eq:SEdfinal}
 R_* = (d-1) \frac{T_{\rm wall}}{V_{\rm eff}(x_i)}, \qquad S_E = \frac{(d-1)^{(d-1)}}{d} V_{S^{d-1}} \frac{T_{\rm wall}^d}{V_{\rm eff}(x_i)^{(d-1)}}\,.
\end{equation}
Again, we note that this approximation is only valid when $V_{\rm eff}(x_i)$ is small and therefore the energy difference between $x_i$ and $x_f$ is small. More precisely, $V_{\rm eff}(x_i)$ must be small compared to the potential bump in between $x_i$ and $x_f$, so that we have $B=S_E\gg1$. Expression (\ref{eq:SEdfinal}) was originally found in \cite{Coleman:1977py,Brown:1988kg} and used in many other papers studying $O(d)$ tunneling events such as \cite{Kachru:2002gs,Klebanov:2010qs}.

\subsection{Particle versus \texorpdfstring{$O(d)$}{} tunneling }\label{ssec:1DvsdD}
We now discuss some notable differences between the tunneling processes and the amplitude $\Gamma = A e^{-B}$ for particles and for extended objects.

\paragraph{Computing $B$ analytically.}
As discussed above, for the particle ($d=1$), the Euclidean action simplifies dramatically: instead of needing to finding the entire trajectory by solving the equations of motion, we can simply integrate the Euclidean momentum $p$ in position space from the starting point $x_i$ to the endpoint $x_f$ of the tunneling process. This is the standard textbook method of calculating tunneling amplitudes in quantum mechanics.

To compute the tunneling parameter $B$ for an $O(d)$ invariant Euclidean bounce, in principle we have to solve the Euler-Lagrange equations subject to the boundary conditions $x(R\rightarrow \infty)=x_i, x(R\rightarrow-\infty)=x_f$. 
These equations can generally only be solved numerically as one needs to know the full trajectory $x(R)$.  An analytic estimate of tunneling rates is possible if the energy of the metastable configuration is small enough. Then the on-shell action is approximated by a contribution of the tension of a thin domain wall and one from the non-zero vacuum energy of the metastable state. 

\paragraph{Computing $A$ and $B$ for metastable and supersymmetric tunneling.}
The behavior of the tunneling coefficient $B$ as a function of the energy difference between the vacua (or the energy of the metastable vacuum $E_{ms} \equiv V_{\rm eff}(x_i)$) can be obtained from  Eq.~(\ref{eq:SEdfinal}) for the $O(d)$-symmetric tunneling and from Eq.~(\ref{eq:pE1D}) for the tunneling of particles. In particular, from (\ref{eq:pE1D}) it is clear that the particle tunneling parameter $B$ can remain finite when $p(x_i)=p(x_f)(=0)$. In contrast, for $O(d)$ tunneling, $B$ blows up\footnote{This is indeed what was found in the various $O(d)$ tunneling models for example in \cite{Kachru:2002gs,Klebanov:2010qs}.} when $E_{ms}\rightarrow0$. We can summarize these observations in the following expansions:
\be \label{eq:BexpansionEms} B_{particle} = B_0 + \mathcal{O}(E_{ms}^1), \qquad B_{O(d)} = B_1 E_{ms}^{-(d-1)} + \mathcal{O}(E_{ms}^{-(d-2)}).\ee
In particular this implies that for the particle, we can simply ignore and set to zero any (small) metastability energy as the finite $E_{ms}$ effects will be subleading. We do exactly that in section \ref{sec:multicenter}.

Note that, while $B$ may remain finite  when $E_{ms}\rightarrow 0$, we do expect the coefficient $A$ in (\ref{eq:Gamma}) to tend to zero in this limit, even for particle tunneling. This is hardly surprising, especially when one uses the standard quantum mechanical picture of waves tunneling through barriers: when the energy of the incoming wave tends to zero, the incoming wave's momentum tends to zero and there is, strictly speaking, no wave left. This means that there is also nothing hitting the barrier and no possibility of a wave exiting the barrier. For example, the transmission coefficient for a rectangular barrier of height $V_0$ at $-a<x<a$ has the small-$E$ expansion (when $a^2mV_0\gg1$):
\be T = \frac{16}{V_0} E \exp(-4a\sqrt{2mV_0}) + \mathcal{O}(E^2,\exp(-8a\sqrt{2mV_0})).\ee
Thus, even though $B$ tends to a finite value as $E\rightarrow 0$, we still have that $A\rightarrow 0$. In section~\ref{sec:multicenter}, we will be only interested in the behavior of the particle tunneling exponent $B$, so we can set $E_{ms}\rightarrow 0$  to calculate $B$ to leading order.

\section{Tunneling of Branes into Black Hole Microstates}\label{sec:multicenter}
We now  discuss the tunneling of branes into smooth, multi-center supergravity backgrounds that can be interpreted as black hole microstate geometries for the three-charge black hole. In M-theory language, the three charges correspond to M2 branes wrapping orthogonal $T^2$ cycles in a compact $T^6$, and in five dimensions these solutions can be described by a $U(1)^3$ ungauged supergravity. Supersymmetry dictates that the five-dimensional metric be a fibration over a four-dimensional hyper-K\"ahler base space \cite{Gutowski:2004yv}. When this base space is of the Gibbons-Hawking form, there is an extra isometry and we can reduce the system along this isometry direction to give a $U(1)^4$ four-dimensional ungauged supergravity theory  known as the STU model. Solutions in this system are determined by eight harmonic functions on $\mathbb{R}^3$ \cite{Bates:2003vx,Gauntlett:2004qy,Bena:2005ni}. In order for the solution to be smooth and devoid of closed time-like curves the locations and residues of the poles 
must satisfy certain particular relations \cite{Denef:2000nb,Bates:2003vx,Bena:2005va,Berglund:2005vb,Bena:2007kg}.

\begin{figure}[ht!]
\centering
\includegraphics[width=0.55\textwidth]{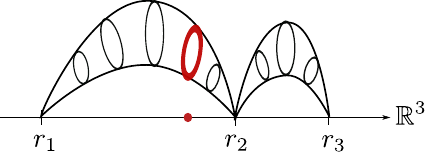}
\caption{Cartoon of a black hole microstate geometry with centers on a line. Depicted in red is a probe brane which in 5D wraps a contractible $S^1$ on the topologically non-trivially two-cycle between two centers and which in 4D reduces to a point particle on a line between the same two centers.}
\label{fig:BHmicro}
\end{figure}

Any probes that we introduce in this system, as long as they have no internal degrees of freedom along the isometry direction of the five-dimensional space, can be described as point particle probes in four dimensions. We use this ``4D/5D connection'' to describe the system in the notation/dimension most convenient to a particular aspect. See Figure~\ref{fig:BHmicro} for an illustration.

In section \ref{ssec:msbackground} and \ref{ssec:msprobe} we use the four-dimensional language of Denef \cite{Denef:2000nb} to calculate the tunneling parameter $B$. We then continue with the five-dimensional conventions of 
\cite{Bena:2007kg}. 
We apply the expression for the tunneling amplitude in section \ref{ssec:mssusy} to calculate the scaling of the tunneling amplitude with the number of centers in two particular choices of background solution. We discuss these results in section \ref{ssec:msimplications} and in particular their implications for the formation of black hole microstates.
To mediate possible confusion we give an extensive review of multi-center solutions in appendix \ref{app:SUGRAconventions} along with an explicit overview of how both sets of conventions are related. More details on calculations of section \ref{ssec:mssusy} are given in appendix \ref{app:multicenter}.

\subsection{Background solution}\label{ssec:msbackground}

We summarize the relevant information on the four-dimensional supersymmetric multi-center solutions of~\cite{Denef:2000nb,Bates:2003vx}. These can for instance be obtained after compactification of IIA string theory on a Calabi-Yau threefold. The electric and magnetic charges of the four-dimensional solutions then correspond to D0, D2, D4 and D6 branes on the internal space, see Table \ref{tab:charges}. We refer to appendix~\ref{app:SUGRAconventions} for more information on the background and the uplift to M-theory with five non-compact directions.

\begin{table}[ht!]
\centering
\begin{tabular}{ccc}
Charge& IIA & M\\
\hline
$\rmp^0$ & D6 & KK monopole\\
$\rmp^A$ & D4 & M5\\
$\rmq_A$ & D2 & M2\\
$\rmq_0$ & D0 & P
\end{tabular}
\caption{Charges and IIA and M-theory interpretation. The charges are defined in terms of the vectors of field strengths $\Gamma = \int dA$}
\label{tab:charges}
\end{table}

The bosonic fields of the solution are the metric, gauge fields $A \equiv (A^0,A^I,A_I,A_0)$ and complex scalars $z^A$. Since we do not need the explicit form of the scalars for the tunneling amplitude, we only present the metric and the gauge fields:
\eal{\label{eq:bosonicfields}
ds^2 &= -e^{2U} (dt+\omega)^2 + e^{-2U} \,d\vec  x  \cdot d\vec  x\,,\\
A &= A_t(dt + \omega) + A_i dx^i\,,
}
where $I=1,\ldots,n_V$ and $n_V$ the total number of vector multiplets\footnote{In four dimensions there are $n_V +1$ vectors (counting the graviphoton). In the specific solutions of section \ref{ssec:mssusy} we have $n_V= 3$, but we can keep $n_V$ generic for the derivation of the tunneling amplitude.}.
The solution is fully determined by a set of functions, conveniently organized into a symplectic vector as $H \equiv (H^0,H^I, H_I,H_0)$, which are harmonic on flat $\mathbb{R}^3$ up to local point sources given by electric and magnetic charges  $\Gamma_i \equiv (\rmp^0,\rmp^I,\rmq_I,\rmp_0)_i$ at positions $\vec r_i$:
\begin{equation}
 H = \rmh + \sum_{i = 1}^N \frac{\Gamma_i}{|\vec r - \vec r_i|}\,,
\end{equation}
with $\rmh = (\rmh^0,\rmh^I,\rmh_I,\rmh_0)$  a vector of integration constants. The explicit solution for the bosonic fields~\eqref{eq:bosonicfields} in terms of these harmonic functions is found by inverting the relation that defines the harmonic functions \cite{Bates:2003vx} in terms of the metric function $U$ and the scalar fields $z^I$:
\be
H \equiv -2 \star_3  d \, [{\rm  Im} \, (e^{U-i \alpha} \Omega(z))]\,,\label{eq:Hdefinition}
\ee
We obtain \cite{Bates:2003vx}:
\begin{equation}
\begin{array}{rlrl}
 e^{2U}&= |Z|\,,\quad  &A_t &=2  {\rm Re}\, (e^{U-i\alpha} \Omega )\,,\\
 \star_3 d \omega &= \langle H,d H\rangle\,, 
 & dA_i \wedge dx^i &= \star_3d H \,.
\end{array}
\end{equation}
In these expressions $\alpha  = {\rm arg}\, Z(\Gamma_{\rm tot})$, with $\Gamma_{\rm tot} \equiv \sum_{i=1}^N \Gamma_i$ the total charge of the solution and the central charge $Z(\Gamma)$ is defined as
\be \label{eq:Zdefinition}
Z(\Gamma) \equiv \langle \Gamma, \Omega(H)\rangle\,, \qquad\text{with}\quad  \langle  \Gamma,\tilde \Gamma\rangle \equiv -\Gamma^0 \tilde \Gamma_0 + \Gamma^I \tilde \Gamma_I -  \Gamma_I \tilde \Gamma^I -\Gamma_0 \tilde \Gamma^0\,.
\ee
$\Omega(H) \equiv \Omega(z(H))$ is the symplectic vector that expresses the dependence of the scalars $z^I$ in terms of the harmonic functions, given in \eqref{eq:Omega}. Again, we do not need its explicit form for the tunneling calculation.

\subsection{Probe branes}\label{ssec:msprobe}

The action for BPS  probe particles with charge $\Gamma$  in a four-dimensional supersymmetric background is:
\be
S = -\frac 1 {G_4}\left[\int  |Z(\Gamma)| ds + \frac 12 \int \langle \Gamma, A \rangle\right]\,,
\ee
where $G_4$ is the 4D Newton constant.
Note the position-dependent mass $m(x) \equiv |Z(\Gamma)|$.
We consider tunneling processes between different centers on a line (see Figure~\ref{tab:charges}), and therefore the path of least resistance is along this line.

The gauge field $A$ only has a time component and an angular component, but no component along the symmetry axis. Hence the assumption of section \ref{ssec:general1D} that $A_i \dot x^i =0$ along the path is justified. Along the path also $\omega =0$ so there are no mixed time-spatial components. Then we can use the results from section \ref{ssec:general1D}  and the definitions of $H$ in \eqref{eq:Hdefinition} and $Z$ in \eqref{eq:Zdefinition} to see that the Euclidean momentum takes the simple expression:
\be\label{eq:pmulticenterIIA}
 |p| =G_4^{-1}e^{-2U} \sqrt{ e^{2U} |Z|^2 - [ {\rm Re} (e^{U-i \alpha} Z(\Gamma) ]^2} = \frac 1{2G_4} |\langle \Gamma, H\rangle|\,.
\ee
The tunneling amplitude is then computed from
\be
B =\frac 1{2 G_4}  \int_{\vec x_i}^{\vec  x_f} |dx| \cdot |\langle \Gamma, H\rangle|\,,
\ee
with $|dx| \equiv \sqrt{d\vec x\cdot d \vec x}$. 

The integrand, the conjugate momentum $|p| = |\langle \Gamma, H\rangle|/2G_4$, has a very natural interpretation. For supersymmetric supergravity configurations with charges $\Gamma_i$ at positions $\vec r_i$, one has that
\begin{equation}
\langle \Gamma_i, H\rangle|_{\vec r = \vec r _i} = \sum_{j\neq i} \frac{\langle \Gamma_i ,\Gamma_j \rangle}{|\vec  r_i - \vec r_j|} + \langle \Gamma_i, h\rangle = 0 \,,
\end{equation}
at every center $\vec r_i$. These equations are obtained as integrability conditions on the rotation one-form $\omega$ and are known as the `Denef equations' or the `bubble equations'. The Denef equations constrain the possible supersymmetric configurations and reduce the $(3N-3)$-dimensional configuration space $\{\vec x_i|i=1\ldots N\}$ for $N$ centers to a $(2N-2)$-dimensional solution space.\footnote{The three centre of mass coordinates are not part of the phase space.} One can also derive these equations by treating each of the centers as a probe in the background sourced by the others. This has been done for supertubes (fluxed D4 branes) in~\cite{Bena:2008dw}.

\subsection{Tunneling process}\label{ssec:mssusy}

We want to model the tunneling of matter undergoing gravitational collapse into black hole microstates. The matter we start from will be a collection of branes of string/M-theory that can tunnel to form a black hole microstate. Since we restrict ourselves to microstates of the three-charge black hole, we start from a collection of branes carrying those three charges. In the remainder of this section we describe the physics in the M-theory frame (for conversion see table \ref{tab:charges}). This means we start from a collection of M2 branes. The end state will be a smooth multi-center supergravity solution or `bubbling solution'. In the M-theory picture, these are made out of several centers carrying Kaluza-Klein monopole charge.

\begin{figure}[ht!]
\centering
\includegraphics[width=.46\textwidth]{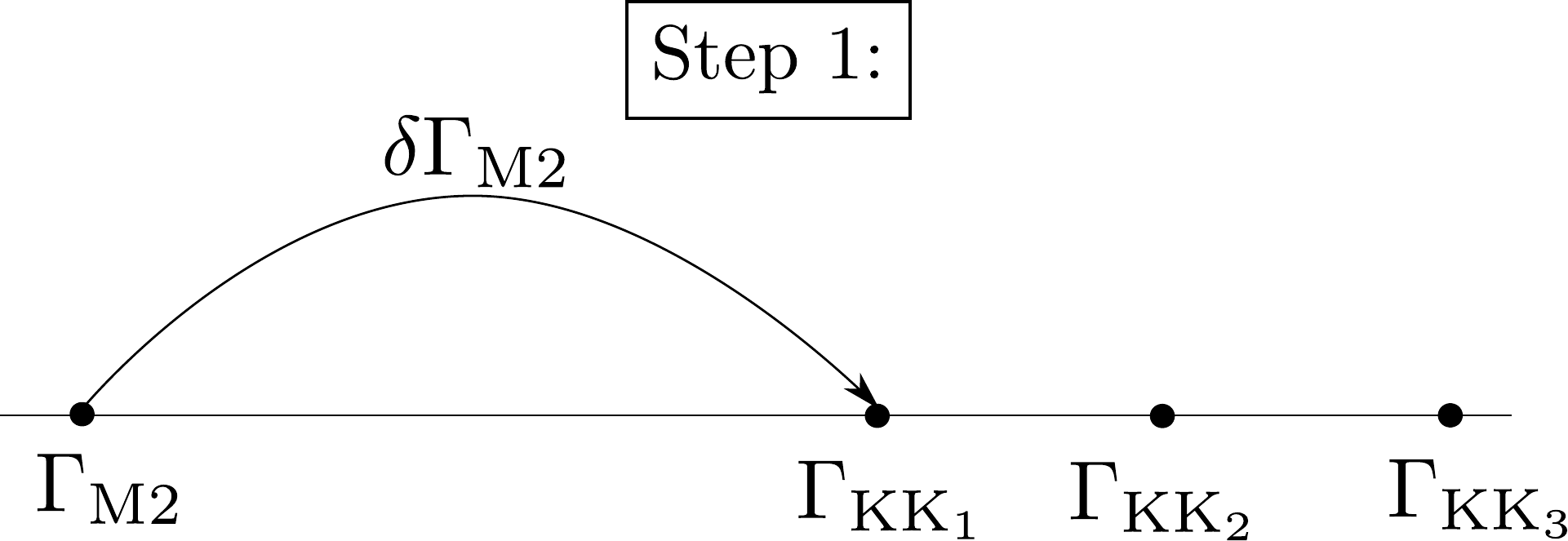}
\hspace{.06\textwidth}
\includegraphics[width=.46\textwidth]{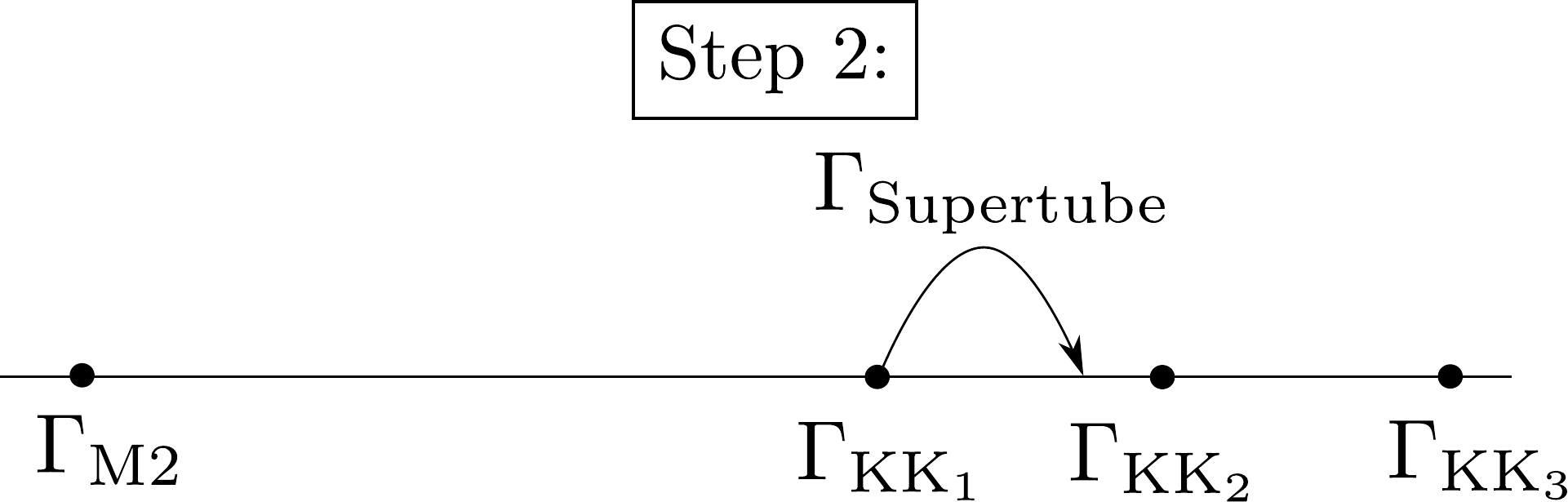}
\caption{Two-step process of bringing M2-charge on top of a KK monopole center and tunneling it into a new supertube center.} \label{fig:steps}
\end{figure}

To enter into the realm of calculations, we imagine a thought experiment in which a very small fraction of the M2 branes have already formed a three-center smooth bubbling solution. This three-center solution can then serve as a catalyst to form more smooth entropy-less centers. To tunnel all the M2 brane charge of the original stack into an $N$-centered microstate geometry, we go through a multi-step process (see Figure \ref{fig:steps}):
\begin{enumerate}[1)]
\item Bring in some of the M2 branes on top of one of the KK monopole centers $\vec r_i$ of the background. 
\item The (now) non-smooth background center $\vec r_i$  can be written as the sum  of  entropy-less constituents. Now `pull away' the extra entropy-less constituents to create a new smooth center (corresponding to a supertube or KK monopole).
\item Continue repeating step (1) and (2) until there are no M2 branes left, but only smooth geometry with a total number of $N$ centers.
\end{enumerate}
The actual  number of centers $N$ and the charge at each center depends on the details of the process. In the examples below we will always divide the M2 brane charge evenly among the newly created centers.

We make some important remarks that can greatly reduce calculational efforts. First, step (1) can be done classically (without the need to tunnel) by allowing the original M2 brane stack to have also momentum charge in the M-theory frame; this step will therefore not reduce the tunneling  rate. This is very natural in an astrophysical setting, as the environment can serve as an angular momentum bath. By adiabatically changing the momentum of the M2 brane stack (exchanging  momentum with the surroundings), one can vary its relative position and freely move the M2 branes over to one of the catalyst centers. 
Note also that the bubble equations of the complete system do not allow to bring all of the M2 brane charge on top of the catalyst in one go. One really needs to repeat the first two steps multiple times to turn all M2-charge into geometry. This is very satisfactory, as at every given moment the small amount of M2-charge brought over to the catalyst immediately tunnels into smooth geometry.

Second, pulling apart the entropy-less constituents in step (2) requires a tunneling process. It is this tunneling rate that we calculate below. For the sake of computation we will not consider the nucleation of new Kaluza-Klein monopole centers. Rather, we will consider entropy-less constituents made out of supertubes. These are the simplest examples of entropy-less  constituents used to build microstate geometries. Moreover, they can easily be considered as probes in multi-center background \cite{Bena:2011fc,Bena:2012zi}. Furthermore, they themselves backreact  to smooth geometry in the D1-D5 frame and, by a spectral flow transformation, one can turn these into Kaluza-Klein monopole centers in the M-theory frame.

In the following we will consider step (2) at an intermediate point in the process. We take an $N$-center bubbling solution, and we imagine tunneling a supertube at the outermost (leftmost) center into a new center further in the interior of the geometry (to the right). We will use specific examples of multi-center solutions and supertube probes to study the process of branes tunneling into topology and flux.
We refer to appendix \ref{app:SUGRAconventions} for more details on both the probe and the background in 4D and 5D and only highlight the main notational conventions here. The relevant parameters of the particular solutions we study are in appendix \ref{app:multicenter}.

We continue with a review of the charge vectors of smooth $N$-center solutions and supertube probes, in both IIA and M-theory frames. Then we calculate the tunneling amplitude for the process explained above for two particular types of solutions. The first type will be a regular (non-scaling) solution and the second will be a particular scaling solution. A multi-center solution is called `scaling' when there exists a limit in which the details of the microstate (charges and fluxes between centers) stay fixed as the centers move arbitrarily close together. The scaling solutions play an important role in the black hole microstate geometry programme, since near the so-called scaling point\footnote{the limit at which the centers all sit on top of each other and the solution has an infinite throat} the solutions are expected to be dual to the typical states of the CFT and thus have a very large entropy \cite{Bena:2006kb,deBoer:2008zn,Bena:2012hf}.

\subsubsection{Supertube probes and smooth microstates}

We first quickly review the notation and interpretation of the charges of the background and probe in four dimensions and their interpretation in the five-dimensional uplift that we will use in the remainder of this paper. The four- and five-dimensional solutions correspond to IIA and M-theory compactifications on a $T^6$, and generalizations to other CY manifolds are straightforward \cite{Cheng:2006yq}.

\paragraph{Four-dimensional solutions and their Type IIA brane interpretation.}

In the four-dimensional description we will focus on the STU model (compactification of IIA string theory on $T^6$) which has $n_V+1=4$ vectors. The background has a total of $2(n_V+1)=8$ charges in the language of section \ref{ssec:msprobe} sourced by D6, D4, D2 and D0 branes. The probes we consider are D4 branes with two lower-dimensional induced D2 charges and D0 charge:
\begin{equation}\label{Gamma4D}
\Gamma \equiv (0;0,0,\rmp^3;\rmq_1,\rmq_2,0;\rmq_1 \rmq_2/\rmp^3)\,. 
\end{equation}
We are interested in tunneling processes of these fluxed D4 branes in $N$-centered backgrounds that are only sourced by D6 and D4 branes. 
At each center $\vec r_i$ the charge vector takes the form
\begin{equation}
 \Gamma_i = \left(\rmp^0_i;\rmp^I_i;\frac{D_{IJK}}{2} \frac{\rmp^J_i \rmp^K_i}{\rmp^0_i};\frac{D_{IJK}}{6}  \frac{\rmp^I \rmp^J_i \rmp^K_i}{(\rmp^0_i)^2}\right)_i\,,
\end{equation}
where $D_{IJK}=|\epsilon_{IJK}|$ and $I=1,2,3$.

\paragraph{Five-dimensional solutions and their M-theory interpretation.}

In five dimensions (obtained for example by compactifying M-theory on a $T^6$) the charges corresponding to D6, D4, D2 and D0 branes become, respectively, Kaluza-Klein monopole or Gibbons-Hawking (GH) charge, magnetic charge (M5), electric charge (M2) and momentum charge (P) (see Table~\ref{tab:charges}). The harmonic functions are written in a different notation as
\begin{equation}\label{eq:4D5DHconvention}
(H^0,H^I,H_I,H_0) \equiv \frac{1}{\sqrt 2} (V,K^I,-L_I,-2M)\,.
\end{equation}
The uplift of the four-dimensional probe with charge vector~\eqref{Gamma4D} becomes a supertube with magnetic dipole charge $d_3$, two electric charges $q_1,q_2$ and a momentum charge $2q_1 q_2/ d_3$ along the wrapped M-theory circle (fifth direction). The  charge vector becomes
\begin{equation}\label{Gamma5D}
\Gamma \equiv \frac 1 {\sqrt{2}} (0;0,0,d_3;-q_1,-q_2,0;-2{q_1  q_2}/{d_3})\,.
\end{equation}
The uplift of the particular choice of IIA background charges turns the D6 branes at $\vec{r}=\vec{r}_i$ into smooth Kaluza-Klein monopoles with GH charge $v_i$ connected by two-cycles threaded by magnetic flux set by flux numbers $k_i$. The charge vector at each center $\vec r_i$ is written as
\begin{equation}
 \Gamma_i= \frac{1}{\sqrt 2}\left(v_i;k^I_i,\frac{D_{IJK}}{2}  \frac{k^J_i k^K_i}{v_i};\frac{D_{IJK}}{6} \frac{k^I_i k^J_i k^K_i}{(v_i)^2}\right)\,,
\end{equation}
Since there are no explicit brane sources the five-dimensional solution is smooth. The magnetic fluxes source total electric charges $Q_I$ and angular momentum $J$ at infinity.

We focus on a background with the $N$ centers on a line.
In this background the Euclidean momentum~\eqref{eq:pmulticenterIIA} becomes
\be \label{eq:pEBW}
|p|= \frac{\pi}{4G_5}\frac{1}{|d_3|}|q_1^{\rm eff} q_2^{\rm eff} V - d_3^2 Z_3|\,,
\ee
with the physical (Maxwell) electric charges of the supertube given by
\be\label{eq:qeff}
q_1^{\rm eff} = q_1 + d_3 \frac{K^2}V\,,\quad q_2^{\rm eff} = q_2 + d_3 \frac{K^1}V\,,
\ee
and where $G_5$ is the 5D Newton constant.
Note that it is the relative sign of the Maxwell charges with respect to the background quantity $Z_3/V$ that determines whether a supertube is BPS or not \cite{Bena:2011fc, Chowdhury:2013pqa}, and not the sign of the Page charges, $ q_1$ and $ q_2$. The latter are oftentimes negative even when the solutions are supersymmetric.

Next we turn to two particular examples. We consider the tunneling of a supertube probe in a non-scaling solution in section \ref{sssec:nonscaling} and in a scaling solution in section \ref{sssec:scaling}. In the rest of this section we will stick to the conventions of \cite{Bena:2007kg}, which are more widely used for five-dimensional microstate geometries.

\subsubsection{Non-scaling solutions}\label{sssec:nonscaling}

\begin{figure}[ht]
\centering
\includegraphics[width=.4\textwidth]{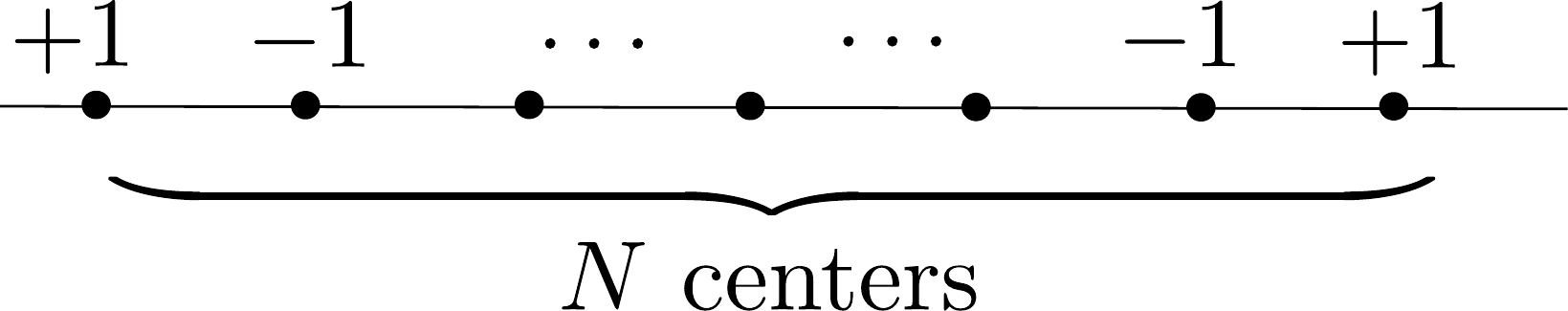}
\caption{Depiction of the $N$-center non-scaling solution with the alternating $+1$'s and $-1$'s corresponding to the GH charges; because $N$ is odd the sum of all these GH charges is $+1$.}
\label{fig:nonscaling}
\end{figure}

As a first example, we take a similar background as studied numerically in \cite{Bena:2006kb}. The background has $N$ centers on a line with $N$ odd; the centers have alternating GH charge $v_j=\pm 1$ such that the total GH charge is unity:  $\sum_j v_j = 1$; see also Figure \ref{fig:nonscaling}. The magnetic fluxes between two centers $i$ and $j$ are
\begin{equation}
 \Pi_{ij}^I = (v_j-v_i) \hat{k}\,,
\end{equation}
where $\hat{k}$ is related to the background dipole charge as $k_i^I=(1-v_i N) \hat{k}$. 
The asymptotic electric charges are
\begin{equation}\label{eq:QIhatkN}
 Q_I = 4 \hat{k}^2 (N^2-1)\,.
\end{equation}

\begin{figure}[ht!]
\centering
\includegraphics[width=0.7\textwidth]{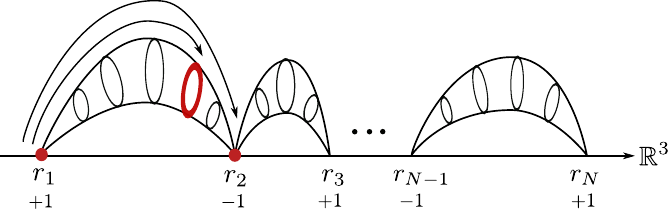}
\caption{Cartoon of a tunneling process in black hole microstates: Starting with a supertube on top of the first center it can tunnel either into a minimum on top of the next center or into a minimum between the two centers. The former process gives a bound on the tunneling amplitude for the latter process.}
\label{fig:BHMStunnel}
\end{figure}

We imagine lowering a supertube probe from infinity to a point on the line near one center, say the left outermost one (see Figure~\ref{fig:BHMStunnel}). In a gauge where there are no Dirac strings on that center we can move the supertube on top of this center; this process does not cost any energy. From this initial position we then compute the amplitude for tunneling into another minimum of the supertube potential.

As discussed in section~\ref{ssec:1DvsdD}, we can compute the tunneling parameter $B$ between initial and final positions where the supertube has supersymmetric minima. Note, however, that in a physical process, to get a non-vanishing coefficient $A$ in the tunneling rate~\eqref{eq:Gamma}, the supertube's initial energy must be above that of its supersymmetric ground state. This can be achieved by starting with a supertube in a metastable minimum close to the center, or, by using a supertube which has some initial kinetic energy. 

When the supertube has tunneled to its final position, wrapping an $S^1$ between the centers 1 and 2 as shown in Figure~\ref{fig:BHMStunnel}, the solution will have a new bubble with flux. This is easiest seen in the duality frame where the charges of the supertube correspond to D1 and D5 branes, in which the common D1-D5 direction shrinks smoothly at the supertube location, giving rise to a nontrivial cycle. Another process also shown in Figure~\ref{fig:BHMStunnel} involves tunneling the supertube all the way from center 1 to center 2. This process is less probable (the supertube has to tunnel through a larger barrier) and will thus give a lower bound on the timescale of a more general tunneling process where the final location is between the two centers. We will focus on this situation here and relegate a more general discussion as well as more information on the background and the tunneling calculation to appendix~\ref{app:morenonscaling}.

The supertube probe has (degenerate) supersymmetric minima on top of a center if one of its physical (Maxwell) charges~\eqref{eq:qeff} vanishes at that center. We choose $q_1$ such that $q_1^{\rm eff}(r_1)=0$ and $q_2$ such that $q_1^{\rm eff}(r_2)=0$. After the tunneling process from center 1 to center 2 the Maxwell charges at center 2 are:
\begin{equation}\label{eq:qeffafter}
q_1^{\rm eff}(r_2) = -2d_3 \hat{k}, \qquad q_2^{\rm eff}(r_2) = 0\,.
\end{equation}

As explained in \cite{Bena:2011fc}, during the tunneling process one needs to keep the Page charge of the supertube constant, but in order to identify the physical charges of the resulting solution one needs to compute its Maxwell charge \footnote{To identify the charge of the supertube one needs to change to a patch where there is no Dirac string wrapped by it, and compute the charge of the resulting configuration. The change of patch alters the Page charge, but leaves the Maxwell charge unchanged. Hence, the Maxwell charge is the physical charge of the supertube solution.}. This is the supertube version of the brane-flux annihilation process \cite{Kachru:2002gs}. For such probes the Euclidean momentum turns out to be \emph{exactly} given by a constant:
\begin{equation}
 |p(r)|=\frac{\pi}{4 G_5} |d_3|\,,
\end{equation}
so that the tunneling parameter $B$ becomes
\begin{equation}
 B=\int_{r_1}^{r_2} dr |p(r)|=\frac{\pi}{4 G_5} |d_3| r_{12}\,,
\end{equation}
where $r_{ij}\equiv |r_i-r_j|$ is the intercenter distance between centers $i,j$. 

To investigate the behavior of $B$ in terms of the physical charges and the number of centers $N$ (keeping $\hat{k}$ fixed), we note that:
\begin{equation}
  d_3 \sim q^{\rm eff}\, \hat{k}^{-1}\,,
\end{equation}
where $q^{\rm eff}$ stands for the physical probe charge that is transfered during the tunneling process, $q^{\rm eff} \equiv q_1^{\rm eff}(r_2)$ see \eqref{eq:qeffafter}. 
One can also show numerically the scaling of the intercenter distance with $N$:
\begin{equation}
r_{12}\sim \hat{k}^2 N^{-\gamma}\,,
\end{equation}
by performing a linear fit on a log-log plot of $N$ versus $r_{12}$. For $N=11+8j$ centers with $j=1,\dots 6$ data points we get $\gamma \approx 1.04$.
Finally, the magnetic flux parameter $\hat{k}$ can be exchanged for asymptotic electric total charges of the geometry $Q\equiv Q_I$ using~\eqref{eq:QIhatkN}:
\begin{equation}
 \hat{k} \sim Q^{1/2}N^{-1}\,.
\end{equation}
This leads to the following scaling:
\begin{equation}
  B \sim q^{\rm eff} \,Q^{1/2} N^{-(1+\gamma)}\,.
\end{equation}

This is how the tunneling amplitude parameter $B$ scales for \emph{one} tunneling event. To estimate the amplitude for tunneling into a black hole microstate of total charge $Q_{tot}$ with $N_{tot}$ centers, we imagine tunneling $N_{tot}$ identical supertube probes with charge $q=Q_{tot}/N_{tot}$. The total tunneling amplitude parameter $B_{tot}$ is then the sum of the parameters $B$ from each of these $N$ tunneling steps. We have, noting that $Q = (Q_{tot}/N_{tot}) N$:
\begin{align} 
B_{tot} &\sim \sum_{N=1}^{N_{tot}} \left(\frac{Q_{tot}}{N_{tot}}\right)^{3/2}N^{1/2-(1+\gamma)}
\label{eq:nonscalingBtotder}\sim Q_{tot}^{3/2} \left(N_{tot}^{-(1+\gamma)}-N_{tot}^{-3/2}\right)\,.
\end{align}
where in the second step we replaced the sum $\sum_{N}$ by the integral $\int dN$. To leading order, we have:\footnote{Note that the minus sign of the $N^{-3/2}$ term in (\ref{eq:nonscalingBtotder}) should not be taken too seriously: since $B= \int |p|$, the end result for $B$ will always be positive; we simply need to choose the overall sign of (\ref{eq:nonscalingBtotder}) to reflect this.}
\begin{equation}
B_{tot} \sim Q_{tot}^{3/2} N_{tot}^{-3/2},
\end{equation} 
Since $\Gamma\sim e^{-B}$, to get the \emph{largest} tunneling probability, we need the \emph{smallest} possible $B$, which implies that tunneling to \emph{more} centers is preferred.

\subsubsection{Scaling solutions}\label{sssec:scaling}

\begin{figure}[ht]
\centering
\includegraphics[width=0.7\textwidth]{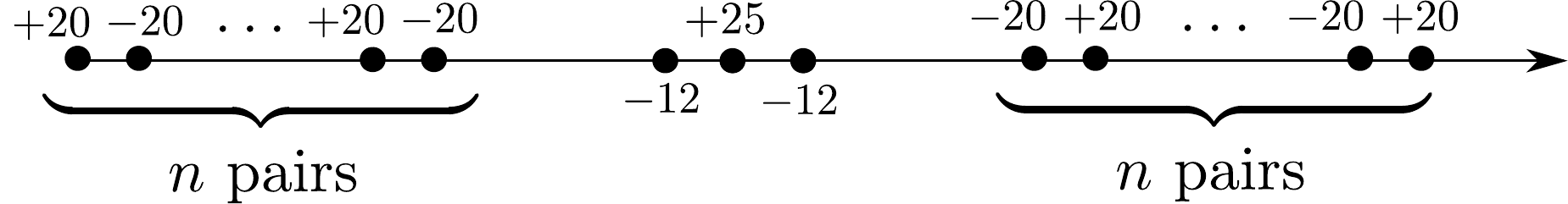}
\caption{Illustration of the $N=4n+3$ center scaling solution. There is a middle triplet with GH charges $-12,+25,-12$ and $n$ pairs on each side of this triplet with GH charges $-20,+20$.}
\label{fig:scaling}
\end{figure}

The second example of a background in which one can study probe supertube tunneling is a scaling solution, inspired by the 7-center scaling solution of \cite{Bena:2006kb,Bena:2012zi}. In this solution, there is a ``middle triplet'' of three centers, and two identical pairs of centers put symmetrically on its sides. We will enlarge this solution by adding a number of identical pairs on the sides (see Figure \ref{fig:scaling}). In this way, we obtain an $N$-centered solution with $N=4n+3$ when there are $n$ extra pairs to each side of the middle triplet. The details of these geometries can be found in appendix \ref{app:morescaling}. The most important feature of these geometries is that the fluxes are determined by a parameter $\hat{k}$. In the limit $\hat{k}\rightarrow k_*$ for a particular numerical value $k_*$ (that depends on the details of all of the background fluxes), the background \emph{scales}: all of the inter-center distances $r_{ij}$ go to zero in this limit, while their ratios remain fixed.

One thing we can immediately see is that in the scaling limit, when $r_{ij}\rightarrow \epsilon\, r_{ij}$ and $\epsilon\rightarrow 0$, all harmonic functions diverge as $1/\epsilon$. This also implies that $q_{1,2}^{\rm eff}\sim \epsilon^0$ since they only involve ratios of harmonic functions. Using (\ref{eq:pEBW}), this simple scaling argument implies that:
\be p \sim \epsilon^{-1},\ee
is the leading behavior, so that the tunneling amplitude for a probe supertube to tunnel from a degenerate supersymmetric minimum on the first, outermost center to the second, adjacent center scales as:
\be B = \int_{r_1}^{r_2} |p| dr \sim\epsilon^0.\ee
In other words, \emph{the tunneling amplitude parameter $B$ does not scale when we scale the distances in the scaling solutions!} This seems to imply that we don't need to consider explicit (fine-tuned) multi-center \emph{scaling} solutions: an analysis for a \emph{non-scaling} solution (such as the above simple alternating solution in section \ref{sssec:nonscaling}) should give qualitatively the same results. We will now verify this with an explicit calculation in our scaling background and compare with the results of section \ref{sssec:nonscaling}.

To find the scaling of $B$ with $q,Q$ and $N$, we must be careful to compare $B$'s between solutions with different $N$ and (approximately) the same asymptotic charges $Q_I$. We should also keep the microstate size fixed which amounts to keeping the distance between the outer centers of the solutions fixed. For the details of this calculation, see appendix \ref{app:morescaling}. The result is:
\be B \sim q^{\rm eff} Q^{1/2} N^{-0.93}.\ee
A similar argument as for the non-scaling solution suggests that to make a black hole microstate of total charge $Q_{tot}$ and centers $N_{tot}$ by tunneling $N_{tot}$ probes with the same charge $q^{\rm eff}=Q_{tot}/N_{tot}$, we should have:
\be B_{tot} \sim Q_{tot}^{3/2} (N_{tot}^{-0.93}-N_{tot}^{-3/2})\sim Q_{tot}^{3/2} N_{tot}^{-0.93}.\ee
Again, since the exponent of $N_{tot}$ is negative, the tunneling into solutions with \emph{more centers} is preferred.

\subsection{Metastable states \& black hole physics}\label{ssec:msimplications}

\subsubsection{Concluding remarks}

For both the scaling and non-scaling solutions, we find that to construct a solution with a (large) number of centers, $N$, and (large) charges $Q\equiv Q_I$, the leading-order behavior of the tunneling exponent is:
\be B \sim Q^{3/2} N^{-\beta},\ee
with $\beta=3/2$ for the non-scaling solutions and $\beta=0.93$ for the scaling solutions.
We note that the $Q$-dependence of $B$ can be determined by dimensional analysis, with the Bekenstein-Hawking entropy $S_{BH}\sim Q^{3/2}$. However, the dependence on $N$ is not. Hence, the fact that the leading-$N$ dependence of $B$ has a negative power ($\beta>0$) in both the scaling and the non-scaling backgrounds is powerful evidence that such a dependence may be a universal feature of the tunneling of branes into any multi-center microstate geometries. It also indicates that, all other things being equal, we will most probably end up in a microstate with more centers. 

As mentioned in section \ref{ssec:1DvsdD}, what we have computed is the leading-order expression for the tunneling parameter $B$ ($B_0$ in (\ref{eq:BexpansionEms})). However, when thinking of an actual tunneling process, we must remember that the tunneling parameter $A$ and thus the tunneling amplitude goes to zero when we take $E_{ms}\rightarrow 0$. Applied to our configuration, this is simply equivalent to the statement that any individual BPS state is stable and will never decay dynamically by tunneling to another BPS state with equal charges. Thus, the statement about the probability to end up in a given microstate should not be taken as a statement about supersymmetric microstates. Instead, we can imagine starting with an initial state consisting of branes which, while close to being supersymmetric, has a (small) extra energy $E_{ms}$. In this pre-microstate, we can imagine $N$ tunneling events resulting in a final microstate geometry with $N$ centers. Because of the extra energy, 
this microstate will not be stationary (its centers could have for example some relative velocities) but will probably decay into a stationary one by the emission of electromagnetic and gravitational radiation. 

Our results give a strong indication that a collapsing shell of (close to supersymmetric) branes will prefer tunneling into a horizonless bubbling solution with as many centers as possible, rather than forming an event horizon. The number of centers of the possible tunneling endpoint is only limited by the quantization of the fluxes wrapping topologically-nontrivial cycles.  However, it is important to realize that we are only considering the tunneling into some very symmetric black hole microstate geometries, and we expect that the states that carry most of the black hole entropy do not have a translational $U(1)$ isometry that allows one to write them using a colinear set of GH centers. For example the superstrata solutions \cite{Bena:2011uw,Giusto:2012jx,Bena:2015bea} that can reproduce the black hole entropy growth \cite{Bena:2014qxa} have naively one or a few bubbles, and are parameterized by arbitrary functions of two variables and hence have no isometry. Similarly, one can obtain a black-hole-like 
entropy from quiver quantum mechanics 
configurations that have three centers \cite{Denef:2007vg}, so it is possible that configurations with a few centers have more entropy.

It is logically possible, but in our opinion very unlikely,  that for black hole microstates with no isometries and very few centers the tunneling parameter $B$ scales differently with $N$. The reason we do not believe this happens is that there is no tunneling barrier that prevents a give supertube to change its shape \cite{Mateos:2001qs,Mateos:2001pi}, neither there is one for a superstratum. Hence the tunneling probability into a wiggly superstratum of the type that carries the black hole entropy will be determined entirely by the topology and fluxes. Since even when $N$ is small the extrapolation of our results still gives a tunneling probability larger than the inverse of the entropy, our calculation establishes that the shell will tunnel with probability one into horizonless structure regardless of the answer to the question whether the black hole entropy is carried by solutions with few or many bubbles. Moreover, if the entropy is carried by solutions with many bubbles, and the number of these bubbles 
scales with the charges, the tunneling probability is parametrically larger than 1, and hence the shell will tunnel into horizonless configurations long before a horizon can form.

\subsubsection{Comparison to Kraus-Mathur}
\label{Kraus-Mathur}

It is important to understand the relation between our calculation and the recent Kraus-Mathur method  \cite{Kraus:2015zda} to estimate the tunneling of a shell. 

The first difference is that our calculation uses explicitly-known black hole microstate solutions, and does not assume any resemblance between these microstate solutions and the black hole solution.
More precisely, the assumption that the tunneling rate of the microstates into a shell of dust is the same as the tunneling of the classical black hole into a shell of dust \cite{Kraus:2015zda} relies on the expectation that the physics of a typical microstate/fuzzball will be reproduced by the black hole solution with a smooth horizon, which is based on the fuzzball complementary arguments of \cite{Mathur:2012jk}. However, the firewall arguments of \cite{Almheiri:2012rt,Almheiri:2013hfa,Avery:2012tf} and the expectations from the explicitly-constructed microstates of \cite{Bena:2012zi} indicate that the observer falling into a typical microstate is more likely to hit a firewall or a solid wall, and will not have a smooth infalling experience of the classical black hole. If these arguments are correct and the typical microstates do not resemble a classical black hole in any way at the scale of the horizon, there is no reason to believe that the tunneling of a classical black 
hole solution into dust and the tunneling of a typical microstate into dust are related, and this would undermine the whole argument of~\cite{Kraus:2015zda}. 

Another assumption made in~\cite{Mathur:2008kg,Kraus:2015zda} is that size of the fuzzballs that describe the typical microstates of the black hole is the same as the size of the black hole. While this may be a reasonable expectation in the framework of the fuzzball proposal, it is not a necessary one. It may be possible, though unlikely, that the geometries dual to typical states have a size that is say 10 times larger than the horizon radius, but their superposition conspires to give a smooth spacetime experience to an infalling observer outside the would-be horizon\footnote{in the same way in which their superposition is argued in the fuzzball complementarity framework to give rise to a smooth spacetime experience for an observer crossing the horizon~\cite{Mathur:2012jk}}.

In contrast, our calculation uses the explicitly-constructed horizonless microstate geometries and does not need to make use of either of these two assumptions.  We find that tunneling of the infalling shell into horizonless microstate geometries happens before the horizon can form, regardless of the size of these microstate geometries. Furthermore, our results are valid without making any assumption on whether the physics of an incoming observer in a fuzzball is approximated by a smooth horizon or by a firewall or by a solid wall. 
Hence, our calculations confirm the robustness of the fuzzball tunneling proposal of~\cite{Mathur:2008kg,Kraus:2015zda,Mathur:2009zs} and indicate that this tunneling will happen regardless of the validity of fuzzball complementarity. 

Besides these differences, there is also a more philosophical difference between the approach \cite{Kraus:2015zda} of calculating the tunneling amplitude of a shell using a black hole solution  and our approach of calculating this amplitude using black hole microstate solutions. The best way to see this is to assume that the fuzzball proposal is {\em not} correct, and therefore all the typical microstates of a black hole have a horizon, and differ from each other by some Planck-scale details near the singularity. The calculation of \cite{Kraus:2015zda} then would imply that the collapsing shell would tunnel with probability one into a solution with a horizon, which is the classical black hole. Hence, this calculation does not establish that a horizon does not form  unless one assumes a-priori that a structure with a right entropy and ``rough'Õ horizon properties can exist there. Another way to say this is that this calculation does not give a direct confirmation of the fuzzball proposal, but only a self-consistency check. On the contrary, our calculation shows that the shell would like to tunnel with probability one precisely into the kind of structure that has been previously constructed to reproduce the black hole entropy, and hence provides a spectacular confirmation of the fuzzball/microstate geometry programme.

\section{Tunneling of Branes in LLM Geometries}\label{sec:LLM}

The 11D LLM bubbling geometries \cite{Lin:2004nb} that are holographically dual to the mass-deformed M2 brane  theory \cite{Bena:2000zb} are in many respects similar to 11D black hole microstate geometries discussed in section~\ref{sec:multicenter}. Both have nontrivial topology threaded by magnetic fluxes, and have no singular sources. All their charges are dissolved into fluxes. Furthermore, both types of geometries allow for metastable M5-M2 brane configurations that can decay to supersymmetric vacua via brane-flux annihilation~\cite{Bena:2011fc,Massai:2014wba}. There are also two key differences. The first is of limited relevance for our calculation: The Killing spinors of the black hole microstate geometries are compatible with M2 branes throughout the solutions, while those of the LLM solution rotate~\cite{Bena:2004jw} from being M2 brane compatible in the asymptotic region to being M5 brane compatible in some intermediate region and anti-M2 compatible in some regions of the deep infrared~\cite{Cheon:
2011gv}.

The second difference is in the calculations of the tunneling rate: The probe branes relevant for tunneling in black hole microstate geometries wrap compact directions but are point-like in the non-compact spacetime. Hence, to study tunneling in black hole microstates, we can simply study the one-dimensional quantum tunneling of an effective particle in the non-compact reduced background, which can be done analytically. On the other hand, the probes relevant for tunneling in LLM geometries are extended along non-compact directions. Hence, to compute tunneling probabilities and lifetimes of metastable states in LLM geometries, we have to compute the full $O(d)$ tunneling amplitude. We will estimate the lifetime of the metastable states found in~\cite{Massai:2014wba} in the approximation where the energy of the metastable state is small, which is a similar approximation to those of~\cite{Kachru:2002gs,Klebanov:2010qs}.

\subsection{Background solution}
\label{ssec:LLMbgd}

There are two ways to study tunneling probabilities for M5 branes with dissolved M2 branes in LLM geometries: either directly via the M5 brane Pasti-Sorokin-Tonin action~\cite{Pasti:1997gx}, or via the Dirac-Born-Infeld--Wess-Zumino action for D4 branes with dissolved F1 strings probing the type IIA reduction of the M-theory solution. The second approach was used in~\cite{Massai:2014wba} to compute the polarization potential. We will use these results to compute the $O(2)$ tunneling action for F1  strings in the IIA background as discussed in section~\ref{sec:basics}. We can then use the result for the decay rate of F1 strings to compute the decay rate of M2 branes in the M-theory background.

\subsubsection{M-theory}
The supergravity solution dual to the mass-deformed M2 brane theory is given by~\cite{Lin:2004nb}:\footnote{As noted in~\cite{Cheon:2011gv,Massai:2014wba}, the four-form field strength~\eqref{11G4} corrects (2.35) of~\cite{Lin:2004nb}.}
\begin{align}
ds_{11}^2 &= H^{-2/3} (-dt^2 + d\omega_1^2+d\omega_2^2) +H^{1/3} \Big[
h^2 (dy^2 + dx^2) + y e^{G} d\Omega_3^2 + y e^{-G}
d\tilde \Omega_3^2\Big]\, , \label{11metric}\\
G_4&= -d(H^{-1} h^{-2} V) \wedge dt \wedge d\omega_1 \wedge d\omega_2 \nonumber \\
&\quad +  \left[ d(y^2 e^{2G} V) - y^3 \star_2 dA \right]
\wedge d\Omega_3+  \left[d(y^2 e^{-2G} V) - y^3 \star_2
  d\tilde A\right] \wedge d\tilde \Omega_3\,, \label{11G4}
\end{align}
with warp factor $H=e^{-2\Phi} = h^2-V^2 h^{-2}$.
The metric describes a three-dimensional external space corresponding to the 
M2 brane worldvolume directions warped on an eight-dimensional transverse 
manifold that consists of a two-dimensional subspace spanned by the 
coordinates $(y,x)$ and two three-spheres $S^3$ and $\tilde{S}^3$. The Hodge star $\star_2$ refers to the flat space spanned by $(y,x)$.
The functions $h,G,V,A, \tilde{A}$ are given in appendix~\ref{app:LLMbackground}.

\subsubsection{IIA theory} 
\begin{figure}[ht!]
\centering
\includegraphics[width=0.65\textwidth]{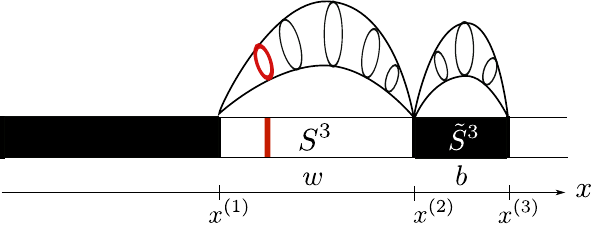}
\caption{Cartoon of an LLM bubbling geometry. Depicted in red is a probe D4 brane (M5 brane) with dissolved F1 strings (M2 branes) wrapping the $S^3$ that has finite size in the white strip region.}
\label{fig:LLM}
\end{figure}

To compute the tunneling rates we will use the IIA reduction along the $\omega_2$ direction\footnote{Upon T-duality along $\omega_1$, this solution is related to the more well-known IIB LLM geometries describing giant gravitons~\cite{Lin:2004nb}.}
\begin{align}
ds^2_{IIA}&= H^{-1} (-dt^2 + d\omega_1^2) + h^2(dy^2 + dx^2) +  y e^{G
} d\Omega_3^2 + y e^{ - G} d \tilde \Omega_3^2\, ,\label{MIIAmetric}\\
B_2 &= -H^{-1} h^{-2} V dt \wedge d\omega_1\, ,\\
F_4 &=  \left[d(y^2 e^{2G} V) - y^3 \star_2 dA\right]
\wedge d\Omega_3 +  \left[d(y^2 e^{-2G} V) - y^3 \star_2
  d\tilde A\right] \wedge d\tilde \Omega_3\label{MIIAF4} \, .
\end{align}
The RR potentials  $C_3=c_3\, d\Omega_3+\tilde c_3 \, d\tilde \Omega_3$ and $C_5=dt\wedge d\omega_1 \wedge(c_5\,  d\Omega_3+\tilde c_5 \, d\tilde \Omega_3)$ can be computed from $F_4=dC_3$ and $\star F_4 =  F_6 =  dC_5 + H_3 \wedge C_3$ and are given explicitly in appendix~\ref{app:LLMbackground}. While all functions in this background depend on both coordinates of the two-dimensional subspace $(y,x)$, the only quantities that enter in the tunneling calculation are the $y\to 0$ limits of these functions. Along the $x$-axis the solutions can then be represented as a sequence of black and white strips, corresponding to different finite-size three-spheres that shrink to zero-size at the strip boundaries, as depicted in Figure~\ref{fig:LLM}.

\subsection{Probe branes}\label{ssec:LLMprobes}

We now apply the results of section~\ref{sec:basics} to the IIA reduced LLM background~\eqref{MIIAmetric}-\eqref{MIIAF4}. See appendix~\ref{app:LLMprobeaction} for details and derivations.
The potential for probe D4 branes with $q$ units of dissolved F1 string charge wrapping the $S^3$ was computed in~\cite{Massai:2014wba}:
\begin{equation}\label{eq:VLLM}
 V_{LLM}=V_{DBI}+V_{WZ}\,,
\end{equation}
with the (rescaled) Dirac-Born-Infeld and Wess-Zumino potentials given by
\begin{equation}\label{eq:DBIWZLLM}
 \hat{V}_{DBI}(x,y)=  H^{-1} \sqrt{H y^3 e^{3G}+(\hat{q}-c_3)^2}\,, \qquad \hat{V}_{WZ}(x,y)=-(\hat{q} B_2 +c_5)\,,
\end{equation}
where we have introduced hats to absorb an overall D4 tension and $S^3$ volume factor: $V_{DBI,WZ}=\mu_4 V_{S^3} \hat{V}_{DBI,WZ}$.\footnote{Throughout this section and appendix~\ref{app:LLM}, the difference between hatted and unhatted quantities will be a factor of the tension of the wrapped brane and the volume of the cycle being wrapped.}
To avoid cumbersome notation we have also introduced a rescaled charge $\hat{q}= q /2\pi \alpha' \mu_4 V_{S^3}$. 
As for black hole microstate geometries, the Maxwell electric charge for F1 strings/M2 branes in LLM geometries given by
\begin{equation}
 \hat{q}^{\rm eff}=\hat{q}-c_3\,,
\end{equation}
determines whether a probe is BPS or not. The key difference here, however, is that both F1 strings and anti-F1 strings (M2 branes and anti-M2 branes) can be compatible with supersymmetry. Conversely, there exist metastable configuration with both signs of the Page charge $q$ depending on the value of $c_3$. We refer to~\cite{Massai:2014wba} and appendix~\ref{app:LLM} for more details. For the purpose of this section we implicitly choose a gauge for $c_3$ in which anti-F1 strings (anti-M2 branes) give rise to metastable states. The decay of these metastable states is then described by the Euclidean bounce action~\eqref{eq:finalBdD}.
Using~\eqref{eq:pEgeneral} and $g_{tt}(x,y)=-H^{-1}$, $g_{xx}(x,y)=h^2$, we obtain the (rescaled) momentum conjugate to $x$:
\begin{equation}
 \hat{p}(x,y) = \sqrt{(\hat{q}-c_3 (1+H^{-1} h^{-2} V^2)+ c_5 V)^2 - h^2 y^3 e^{3G} (1-y e^G)}\,,
\end{equation}
We will only be concerned with the $y\to 0$ limit where the potential \eqref{eq:VLLM} is smallest. In a patch with no Dirac strings at the boundary $x^{(i)}$, given by $c_3(x^{(i)})=0$, the (rescaled) momentum becomes:
\begin{equation}
\label{eq:pELLM}
 \hat{p}(x)=|\hat{q}-2 (x-x^{(i)})|\,.
\end{equation}
The tunneling amplitude parameter is then computed from
\begin{equation}
 B=\int_{S^{1}} d\Omega_{1} \int_{R_i}^{R_f} dR R \,\partial_R x \,p(x)\,.	 
\end{equation}

\subsection{Bubble nucleation} \label{ssec:LLMtunnel}
We now apply the tunneling results of section~\ref{sec:basics} to compute the decay of metastable probe branes in LLM geometries. As discussed, we can obtain an analytic estimate of the lifetime of the metastable states when the energy above the supersymmetric vacuum state is small.

\subsubsection{The $O(2)$-invariant action describing the tunneling of F1 strings in IIA}

We calculate the tunneling of F1 strings extended along $t,\omega_1$ into F1 strings polarized into D4 branes wrapping an additional $S^3$ at the position $x$. The Euclidean radial coordinate is $R=\sqrt{t+\omega_1^2}$. The decay of the metastable vacuum at $x_i$ to the supersymmetric one at $x_f$ is described by the trajectory $x(R)$.
The Euclidean action per unit $\omega_1$ length is
\begin{equation}
 S_E = V_{S^1} R_* T_{\rm wall} - \frac{V_{S^1}}{2} R^2_* V_{\rm eff}(x_i) \,,
\end{equation}
with $V_{S^1}=2\pi$.
The solution to the Euclidean equations of motion minimizes the action, $dS_E/dR_*=0$, giving
\begin{equation}
 R_* = \frac{T_{\rm wall}}{V_{\rm eff}(x_i)}\quad \Rightarrow \quad S_E =  
 \frac{\pi T_{\rm wall}^2}{V_{\rm eff}(x_i)}\,.
\end{equation}
The decay rate has two contributions: one coming from the tension of the domain wall and one from the non-zero vacuum energy of the non-supersymmetric state.

\paragraph{Domain wall.}
The domain wall is described by the trajectory $x(\omega_1)$, where $\omega_1$ is transverse to the wall that interpolates between the vacua $x_i$ and $x_f$.
The Euclidean action is
\begin{equation}
 S_E=  \int dt \int d\omega_1 L_E = \int dt \int d\omega_1  \partial_1 x\, p(x)=  \int dt \int_{x_i}^{x_f} dx \, p(x)\,,
\end{equation}
where in the last equality we used the fact that $H_E=p \partial_1 x -L_E=0$ and rewrote the integral as an integral over $x$. 
We have to integrate~\eqref{eq:pELLM} from $x_i \approx x^{(i)}$ to $x_f=x^{(i+1)}=x^{(i)}+w$ where $w$ is the width of the white strip corresponding to the wrapped $S^3$. 

The (rescaled) tension of the domain wall is given by
\begin{equation}\label{eq:Twall}
 \hat{T}_{\rm wall}=\int_{x_f}^{x_{i}} dx\,\hat{p}(x)
 \approx w^2\,.
\end{equation}
Restoring the mass dimension $\mu$ in the flux in~\eqref{11G4} and \eqref{MIIAF4} as in~\cite{Cheon:2011gv}, we have $w\sim \mu$. 
The quadratic dependence of the domain wall tension~\eqref{eq:Twall} on the mass deformation  agrees with~\cite{Bena:2000zb} and with the field theory result~\cite{Hanaki:2008cu}. Explicitly, the tension of the domain wall obtained from the D4 brane wrapped on the $S^3$ is $T_{\rm wall}^{D4}=\mu_4 V_{S^3} w^2$.

\paragraph{Metastable state.}
The (rescaled) effective potential that characterizes the metastable state when the energy above the supersymmetric vacuum state is small is given by~\cite{Massai:2014wba}:
 \begin{equation}
 \hat{V}_{\rm eff}(x)=
 \hat{q} (a_1 +a_2 x) -\frac{4}{\hat{q}} x^3\,, 
 \end{equation}
where $a_1,a_2$ are defined by
\begin{equation}
 -\lim_{y\to0}\left[ H^{-1}+B_{t\omega_1}\right] = a_1 +a_2 (x-x^{(i)}) + \mathcal{O}\left( (x-x^{(i)})^2\right)\,.
\end{equation}
The leading behavior of the effective potential at the metastable minimum $x_i$ is determined by the energy of the metastable anti-F1 strings at the boundary $x^{(i)}$:
\begin{equation}
 V_{\rm eff}(x_i) \approx V_{\rm eff}(x^{(i)}) = \frac{|q|}{2\pi \alpha'} \lim_{y\to0}\left[ H^{-1}+B_{t\omega_1}\right]\Big|_{x=x^{(i)}} 
 \equiv  |q| V_{F1}(x^{(i)})\,,
\end{equation}
where we defined the energy for one such anti-F1 string ($|q|=1$). See appendix~\ref{app:LLMprobeaction} for more details.\\

Putting it all together, tunneling from a metastable minimum close to the boundary $x^{(i)}$ to the supersymmetric minimum at the boundary $x^{(i+1)}=x^{(i)}+w$ is given by
\begin{equation}
B\approx \frac{\pi (T_{\rm wall}^{D4})^2}{|q| V_{F1}(x^{(i)})}\equiv \alpha_{IIA} \frac{\mu^4}{|q|}\,.
\end{equation}
The tunneling rate $\Gamma\sim e^{-B}$ is thus suppressed for small charge $|q|$ and for large mass deformation $\mu$.

\subsubsection{The \texorpdfstring{$O(3)$}{}-invariant action describing the tunneling of M2 branes}

What happens if we do not compactify on $\omega_2$? Strictly speaking we would have to start from the M5 brane action with dissolved M2 charge. However, in the limit of small $\omega_2$ this action should reduce to that of a D4 brane with dissolved F1 charge. The tension of the domain wall is approximated by the BPS M5 brane wall obtained as the uplift of the IIA construction. Similarly, the energy of anti-M2 branes at the odd boundary $x^{(i)}$ close to the metastable minimum is obtained from the uplift of the anti-F1 strings above. So, we have
\begin{equation}
 V_{M2} = \frac{V_{F1}}{2\pi R_{11}}\,, \quad T_{\rm wall}^{M5}=\frac{T_{\rm wall}^{D4}}{2\pi R_{11}}\,,
\end{equation}
where $\alpha' R_{11}= g_s l_s^3 = l_P$.
The contribution to the bounce action is then obtained from the Euclidean action per unit $\omega_1-\omega_2$ area
\begin{equation}
 S_E= V_{S^2} R_*^2 T_{\rm wall}^{M5} -\frac{V_{S^2} }{3}R_*^3 V_{\rm eff}^{M2}\,,
\end{equation}
with $V_{S^2}=4\pi$, which is extremized at
\begin{equation}
 R_*=\frac{2 T_{\rm wall}^{M5}}{V_{\rm eff}^{M2}} \quad \Rightarrow \quad S_E=\frac{16 \pi (T_{\rm wall}^{M5})^3}{3 (V_{\rm eff}^{M2})^2}\,.
\end{equation}
The tunneling action from a metastable minimum close to the boundary $x^{(i)}$ to the supersymmetric minimum at the boundary $x^{(i+1)}=x^{(i)}+w$ is given by
\begin{equation}
B \approx \frac{16 \pi (T_{\rm wall}^{M5})^3}{3 |q|(V_{M2}(x^{(i)}))^2} \equiv \alpha_{M} \frac{\mu^6}{|q|^2}\,.
\end{equation}
The tunneling rate is again suppressed for small charge $|q|$ and large mass deformation $\mu$. Note, however, that the scaling of the tunneling parameter $B$ with $q$ and $\mu$ in M-theory is different from tunneling in the IIA reduced background. The reason is that for the $O(2)$ case we compute tunneling per unit length while in the $O(3)$ case we compute tunneling per unit area.

\subsection{Metastable states \texorpdfstring{$\&$}{} mass-deformed M2 brane theories}

For both the 11D LLM geometries and their IIA reduction we computed the leading-order behavior of the tunneling exponents in the approximation where the the metastable state has only small excess energy compared to the supersymmetric vacuum. The scaling of the tunneling exponents with the charge $q$ and mass deformation parameter $\mu$ can be summarized as:
\begin{equation}
 B \sim \mu^{2d}/|q|^{d-1} \,,
\end{equation}
where $d=3$ for the 11D solutions and $d=2$ for the 10D solutions. The decay of metastable anti-M2 branes or anti-F1 strings in the infrared of LLM geometries is thus highly suppressed for small charge $|q|$. In~\cite{Massai:2014wba} it was suggested that these long-lived configurations should, via the gauge/gravity duality, correspond to metastable states in the mass-deformed M2 brane theory~\cite{Bena:2000zb}. 

There is one other example where a gravity dual was used to conjecture the existence of metastable states in a strongly-coupled 2+1 dimensional  theory - in \cite{Klebanov:2010qs} it was shown that probe M2 branes in the CGLP solution~\cite{Cvetic:2000db} can have metastable minima. However, at this point there is no technology that may allow one to look for the existence of such minima in the dual theory. Such a technology only exists for 3+1 dimensional theories~\cite{Intriligator:2006dd}, and one may hope that the recent progress in finding new dualities in three dimensions \cite{Aharony:2013dha} may allow for the development of such a technology for searching metastable vacua of CGLP and mass-deformed M2 brane theories.

\section*{Acknowledgments}
We would like to thank Jan de Boer, Freddy Cachazo, Juan Maldacena, Stefano Massai, Thomas Van Riet and Nick Warner for enlightening discussions. 
The work of I.B. was supported in part by the ERC Starting Grant 240210 String-QCD-BH, by the John Templeton Foundation Grant 48222 and by a grant from the Foundational Questions Institute (FQXi) Fund, a donor advised fund of the Silicon Valley Community Foundation on the basis of proposal FQXi-RFP3-1321 (this grant was administered by Theiss Research). D.R.M is supported in part by NSF CAREER Grant PHY-0953232. This work is part of the research programme of the Foundation for Fundamental Research on Matter (FOM), which is part of the Netherlands Organisation for Scientific Research (NWO). 
The work of A.P. is supported by National Science Foundation Grant No. PHY12-05500.
B.V. is supported by the European Commission through the Marie Curie Intra-European fellowship 328652--QM--sing.
We are grateful to the Centro de Ciencias de Benasque Pedro Pascual for hospitality during this work.

\appendix{

\section{\texorpdfstring{$\mathcal{N}=2$}{} Supergravity in Four and Five Dimensions}\label{app:SUGRAconventions}

We compare the common conventions of the four-dimensional multi-center solutions and those of five-dimensional bubbling solutions. We discuss the action, the fields, the multi-center solutions and the probe potential in both conventions. The reader with only an interest in the matching of the two sets of conventions can skip to section \ref{ssec:matching}

\subsection{Four dimensions}

\subsubsection{$\mathcal{N}=2$ supergravity in 4 dimensions}

We consider $N=2$ supergravity coupled to $n_V$ vector multiplets.  There are $n_V$ scalars and $n_V+1$  vector fields, and the action is
\begin{align}
S_4 =  \frac1{16 \pi G_4}\int \Big{(}& \star_4 R_4-2 g_{I \bar J} \star_4 d z^I \wedge d \bar z^{\bar J}\\
&- c^2 ({\rm Im\,\caln}_{IJ}) \star_4 F^\Sigma \wedge F^\Lambda -c^2 ({\rm Re}\, \caln_{\Sigma \Lambda}) F^\Sigma\wedge F^\Lambda\Big{)},
\label{eq:4dAction}
\end{align}
with the index $I = 1 \ldots n_V$, $F^\Sigma = (F^0,F^I)$, and $c$ a convention-dependent constant. For instance Denef \cite{Denef:2000nb} takes $c = 1$. The scalar metric $g_{I \bar J} $ and period matrix $\cal N$ depend on the scalars $z^I$ and their complex conjugates.

The couplings of the theory are determined from one single function, a holomorphic prepotential $F(X)$, that is holomorphic of degree two in the projective coordinates $(X^0(z),X^I(z))$. The functions $g_{I\bar J} (z), {\cal N}_{\Sigma\Lambda}(z)$ appearing in the action are most conveniently written by introducing the symplectic section
\be
\Omega_0 = (X^\Sigma, F_\Sigma)\,,\qquad F_\Sigma = \frac{\partial F}{\partial X^\Sigma}\,,
\ee
which is endowed with a natural symplectic product
\begin{equation}
\langle\Omega_0,\tilde \Omega_0\rangle = -X^0\tilde X_0 + X^I\tilde F_I - F_I\tilde X^I + F_0 \tilde X^0 \,.\label{eq:sympproduct}
\end{equation}
The metric is special K\"ahler, as it can be obtained from a K\"ahler potential $\cal K$ derived from the prepotential: 
\be
g_{I\bar J} = \partial_I \bar \partial_{\bar J} {\cal K}\,, \qquad {\cal K} = - \ln i \langle \Omega_0, \bar  \Omega_0\rangle\,,\qquad z^I  =  \frac{X^I}{X^0}
\ee
We can and will set $X^0=1$. The vector kinetic couplings are
\be
\caln_{\Lambda\Sigma} = \bar F_{\Lambda\Sigma} + 2 i \frac{({\rm Im}\, F_{\Lambda\Omega}X^\Omega)( {\rm Im}\, F_{\Sigma\Omega'}X^{\Omega'})}{F_{\Omega\Omega}X^\Omega X^{\Omega'}}\,,\qquad F_{\Sigma \Lambda} = \frac{\partial^2 F}{\partial X^\Sigma \partial X^\Lambda}\,.
\ee

Finally the gauge fields are determined by the electric and magnetic charges. We define the dual field strengths and potentials as
\begin{equation}
G_\Sigma \equiv d A_\Sigma \equiv {\rm Re}\, {\cal N}_{IJ} d A^J + {\rm Im}\, {\cal N}_{IJ} \star_3 d A^J\,.\label{eq:DualGaugeFieldsDef}
\end{equation}
The charges are
\be
p^\Sigma = \frac 1 {4\pi} \int    F^\Sigma  \,, \qquad q_\Sigma = \frac 1 {4\pi} \int   \star_4 G_\Sigma  
\ee
We will package these in a symplectic charge vector
\begin{equation}
\Gamma \equiv (p^\Sigma,q_\Sigma)\,.
\end{equation}
which also has a natural symplectic product \eqref{eq:sympproduct}

For later use we also define the non-holomorphic symplectic section
\begin{equation}
 \Omega \equiv e^{{\cal K}/2} \Omega_0\label{eq:Omega}\,.
\end{equation}

One can obtain the action \eqref{eq:4dAction}from a Calabi-Yau compactification of type II supergravity. Then the prepotential is
\be
F  = \frac 16 D_{IJK}\frac{ X^I X^J X^K}{(X^0)^2}\,,
\ee
with $D_{IJK}$ triple intersection numbers. 
We are interested in a IIA compactification. Then the charges of the solution correspond to wrapped D6, D4, D2 and D0 branes, as in Table \ref{tab:charges}. They are related to integer flux numbers $N^I,N_I$ as
\begin{equation}
p^\Sigma = \frac{\sqrt{8} }{c} T^\Sigma V^\Sigma G_4 N^\Sigma\,,\qquad q_\Sigma = \frac{\sqrt{8} }{c} T_\Sigma V_\Sigma G_4 N_\Sigma\,,
\end{equation}
with $T^\Sigma,T_\Sigma$ the tensions of the D-branes in Table \ref{tab:charges}, and $V^\Sigma,V_\Sigma$ the volumes of the cycles they are wrapping.

\subsubsection{Probe particles}

The probe potential for a BPS particle with charges $\Gamma$ in a supersymmetric background solution to the $N=2$ action is 
\be
S = -\int  |Z(\Gamma,t)| ds - \frac 12 \int \langle \Gamma, A_\mu\rangle \frac{dx^\mu}{ds} ds\,.\label{eq:probe_denef}
\ee
The function $Z$ is known as the central charge:
\be
Z(\Gamma,t) = \frac 1 {\sqrt{\frac 4 3 ({\rm Im}\, z)^3}}\left(\frac 16 \Gamma^0 z^3 - \frac 12 \Gamma^A z_A + \Gamma_A z^A - \Gamma_0\right)\,,\label{eq:centralcharge}
\ee
with the definitions
\be
z_A^2 \equiv D_{ABC} z^B z^C\,,\qquad z^3 \equiv D_{ABC} z^A z^B z^C\,,\qquad ({\rm Im}\, z)^3 \equiv D_{ABC}({\rm Im}\, z^A)({\rm Im}\, z^B)({\rm Im}\, z^C)\,.
\ee
The potential has supersymmetric minima at positions in $\mathbb{R}^3$  for which
\begin{equation}
 \langle \Gamma, H\rangle = 0\,.
\end{equation}

\subsubsection{BPS multicenter supergravity solutions}\label{ss:background}

The most general multi-center supersymmetric background  of $N=2$ supergravity is fully determined by a symplectic vector of harmonic functions \cite{Denef:2000nb,LopesCardoso:2000qm,Bates:2003vx}
\begin{equation}
 H = (H^0,H^I,H_I,H_0)
\end{equation}
trough a single function $\Sigma$ that is homogeneous of degree two:
\begin{eqnarray}
ds^2 &=& -\Sigma(H)^{-1} (dt+\omega)^2 + \Sigma(H) \,dx^{i} dx^{i}\,,\\
t^A &=& \frac{H^A-i\frac{\partial \Sigma}{\partial H_A}}{H^0+i\frac{\partial \Sigma}{\partial H_0}}\,,\\
A &=& I\cdot\frac{\partial \log \Sigma(H)}{\partial H} (dt + \omega)+ A_i  dx^i\,,
\end{eqnarray}
where $I$ is the symplectic matrix that defined the symplectic product:
\begin{equation}
I = \begin{pmatrix}0&0&0&-1\\0&0&\unity&0\\0&-\unity&0&0\\1&0&0&0\end{pmatrix}\,.
\end{equation}
The spatial  one-forms forms are defined as
\begin{eqnarray}
\star_3 d (A_i dx^i) &=& dH\,,\\
\star_3 d \omega &=& \langle dH, H\rangle\,.
\end{eqnarray}
The integrability condition $d\star_3  d \omega  = 0$ leads to $\langle \Gamma_i, H\rangle =0$ for every charge vector $\Gamma_i$ at position $\vec r_i$. With harmonic functions $H  = h + \sum_i \Gamma_i/|\vec r - \vec r_i|$, this can  be written as the set of Denef  equations:
\begin{equation}
\sum_{j\neq i} \frac{\langle \Gamma_i , \Gamma_j\rangle}{|\vec r_i -\vec r_j|} = \langle h, \Gamma_i\rangle\,.
\end{equation}
For $N$ centers, there are $N-1$ independent Denef equations.

The function $\Sigma (H)$ is determined from 
\begin{eqnarray}
\Sigma &=& \sqrt{\frac{Q^3 - L^2}{(H^0)^2}}\,,\nonumber\\
L &=& H_0 (H^0)^2 + \frac 13 D_{ABC}H^A H^B H^C-H^A H_A H^0\,,\nonumber\\
Q^{3/2} &=&\frac 13 D_{ABC} y^A y^B y^C\,,\nonumber\\
D_{ABC} y^B y^C &\equiv& - 2 H_A H^0 + D_{ABC}H^B H^C\,.
\end{eqnarray}


The explicit solutions for the scalars and the time components of the vectors are
\begin{eqnarray}
t^A &=& \frac{Q^{3/2}H^A -L y^A}{H^0 Q^{3/2}} + i \frac{\Sigma y^A}{Q^{3/2}}\,,\label{eq:Moduli_MultiCenter}\\
A^0 &=&\frac{L }{\Sigma^2}(dt + \omega) +A_d^0\,,\nonumber\\
A^A &=& \frac{H^A L- y^A Q^{3/2} }{\Sigma^2 H^0}(dt + \omega)+A_d^A\,,\nonumber\\
A_A &=& \frac{L (-H_A H^0+D_{ABC} H^B H^C )- D_{ABC} y^B H^C Q^{3/2}}{\Sigma^2(H^0)^2}(dt + \omega)+(A_d)_A\,,\nonumber\\
A_0 &=& \left(\frac{L ( H^A H_A-2 H_0 H^0)- y^A H_A Q^{3/2}}{\Sigma^2(H^0)^2}
-  \frac{d H^0}{H^0}\right)(dt + \omega)+(A_d)_0\,.
\end{eqnarray}

\subsection{Five dimensions}

We are only  concerned with five-dimensional $N=2$ supergravity with two vector multiplets, that reduces to the four-dimensional STU model.

\subsubsection{STU model in five dimensions}

The action of 5d minimal supergravity with 2 vector multiplets is (we use the conventions of \cite{Bena:2007kg}):
\be
S_5 = \frac 1{16 \pi G_5}\int \left(\star_5 R_5 - Q_{IJ} \star_5 d y^I \wedge d y^J - Q_{IJ} \star_5 \tilde F^I \wedge \tilde F^J - \frac 1 6 C_{IJK} \tilde F^I \wedge \tilde F^J \wedge \tilde A^K\right)\,,\label{eq:5dAction}
\ee
where now $I = 1,2,3$. We put tildes on the five-dimensional vector field to avoid confusion with the four-dimensional ones. The vector multiplet kinetic matrix is
\be
Q_{IJ} = \frac 12 (y^I)^{-2}{\delta_{IJ}}
\ee
The scalars obey the restriction
\be
\frac 1 6 C_{IJK} y^I y^J y^K = 1\,,
\ee
with
\be
C_{IJK} = |\epsilon_{IJK}|\,.
\ee

\subsubsection{Multi-center solutions in five dimensions}

The metric, scalars and gauge fields of supersymmetric solutions with a timelike Killing vector have the form  \cite{Bena:2004de,Gutowski:2004yv}:
\bea
ds_{5}^2 &=& -(Z_1 Z_2 Z_3)^{-2/3}(dt + k)^2 + (Z_1 Z_2 Z_3)^{1/3}\,ds_4^2\,,\\
y^I &=& \frac{(Z_1 Z_2 Z_3)^{1/3}}{Z_I}\,,\\
A^I &=& \left(-Z_I^{-1}{(dt + k)} + B^{I}\right)\,,\label{eq:5d_solution}
\eea
where $ds_4^2$ is a four-dimensional hyper-K\"{a}hler metric. The rotation one-form $k$ and the magnetic potentials $B^{I}$ are supported and only depend on this four-dimensional base space and also the warp factors $Z_I$ only depend on the coordinates of the base. 

We focus on solutions where the 4d base is Gibbons-Hawking:
\be
d s^2_4 = V^{-1} (d \psi + A)^2 + V ds^2_3 (\mathbb{R}^3)\,,\qquad \star_3 d A = - d V,
\ee
with $V$  a harmonic function on $\mathbb{R}^3$. Solutions with a GH base have a natural interpretation upon KK reduction along the GH fibre $\psi$ as four-dimensional multi-center solutions.

A generic multi-center supersymmetric 3-charge solution in five dimensions is determined by 8 harmonic functions $(V,K_I,L^I,M)$  on $\mathbb{R}^3$ \cite{Gauntlett:2004qy,Elvang:2004ds}. The warp factors $Z_I$, magnetic fields $B^I$ and angular momentum one-form are
\begin{align}
B^{I} &= V^{-1} K^I (d \psi + A) +\xi^I,\qquad d \xi^I = - \star_3 d K^I\nonumber\\
Z_I &= L_I + \tfrac 12  D_{IJK}V^{-1} K^J K^K \nonumber\\
k &= \mu (d \psi + A) + \omega\,,
\end{align}
with
\begin{align}
\mu &= \tfrac 16 V^{-2} C_{IJK} K^IK^JK^K + \tfrac12 V^{-1} K^I L_I + M\,,\nonumber\\
\star_3 d \omega &= V d M - M d V + \frac 12 (K^I dL_I - L_I d K^I)\,.
\end{align}

\subsubsection{Probe  supertubes}

The potential for a probe supertube with 5D dipole charge $d_3$ and two  electric charges $q_1,q_2$ in a background \ref{eq:5d_solution} is \cite{Bena:2011fc}:
\begin{align}
 V =\,& \frac  1 {d_3}\frac{\sqrt{Z^3/V}}{Z^3/V-\mu^2}\sqrt{\tilde q_1^2 + d_3^2\frac{Z^3/V - \mu^2}{Z_2^2} }\sqrt{\tilde q_2^2 + d_3^2\frac{Z^3/V - \mu^2}{Z_1^2} }\nonumber\\
  &+ \frac 1 {d_3}\frac{\mu \tilde q_1 \tilde  q_2}{Z^3/V-\mu^2}  - \frac{\tilde q_1}{Z_1} -\frac{\tilde q_2}{Z_2} - d_3\frac{\mu}{Z_1  Z_2}\,,\label{eq:probe_iosif}
\end{align}
with the definitions
\begin{equation}
 \tilde q_1  \equiv q_1 + d_3 \left(\frac{Z_2}V-\frac {\mu}{Z_2}\right) \,,\qquad \tilde q_2 \equiv q_2 + d_3  \left(\frac{Z_1}V-\frac {\mu}{Z_1}\right)\,.
\end{equation}
The supertube potential has supersymmetric minima at positions in $\mathbb{R}^3$ for which
\begin{equation}
 \frac{Z_3}V = \frac{(q_1  + d_3  Z_2/V )( q_2+ d_3 Z_1/V)}{(d_3)^2}\,.
\end{equation}
This equation is sometimes referred to as the supertube radius relation.

\subsubsection{Reduction to four dimensions}

We reduce the action \eqref{eq:5dAction} to four dimensions with the ansatz:
\bea
ds_5^2 &=& e^{2\beta \varphi} (dx^5 - A^0) + e^{-\beta\varphi} ds_4^2\,,\nonumber\\
\tilde F^I &=& d(a^I(dx^5 - A^0) + A^I)\nonumber\\
&=& da^I \wedge (dx^5 - A^0) + F^I - a^I F^0\,.
\eea
Up to total derivatives, this gives the action in four dimensions:
\begin{align}
\nonumber
S_4 = \frac1{16\pi  G_4}\int \Big{(}&\star_4 R_4 -\frac{3}{2} \beta^2\star_4 d \varphi \wedge  d \varphi- Q_{IJ} \star_4 d y^I \wedge d y^J - Q_{IJ} e^{-2\beta\varphi}\star_4 d a^I \wedge d a^J\nonumber\\
&-e^{3\beta\varphi} \star_4 F^0\wedge F^0-Q_{IJ} e^{\beta \varphi}\star_4 (F^I - a^I F^0) \wedge ( F^J - a^J F^0)\\
&-\frac 12 C_{IJK} a^K F^I \wedge F^J+\frac 12 C_{IJK} a^J a^K F^I\wedge F^0-\frac 16 C_{IJK} a^I a^J a^K F^0\wedge F^0\Big{)}\,.
\nonumber
\end{align}
with 
\be
G_4 = G_5/(2\pi L_5)
\ee
and $L_5$ is the length of the $x^5$ circle. We rewrite the action to comply with the notation of \eqref{eq:4dAction}:
\begin{align}
S_4 =  \frac1{16\pi  G_4}\int \Big{(}& \star_4 R_4-\frac 12 \sum_{I=1}^3 \frac{\star d z^I \wedge d \bar z^I}{({\rm Im}\,z^I)^2}\\
&- \frac 12 ({\rm Im\,\caln}_{\Lambda\Sigma}) \star_4 F^\Sigma \wedge F^\Lambda - \frac 12 ({\rm Re}\, \caln_{\Lambda\Sigma}) F^\Sigma\wedge F^\Lambda\Big{)}.
\end{align}
The complex scalar fields $z^I$ are related to the real scalars $y^I, \varphi, a^I$ as:
\begin{align}
z^I \equiv a^I + i b^I \equiv a^I + i e^{\beta \varphi} y^I\,.
\end{align}
and the matrices in this action are
\begin{align}
{\rm Im}\,\cal N&=b_1 b_2 b_3\left(
\begin{array}{cccc}
 1+\frac{a_1^2}{b_1^2}+\frac{a_2^2}{b_2^2}+\frac{a_3^2}{b_3^2} & -\frac{a_1}{b_1^2} & -\frac{a_2}{b_2^2} & -\frac{a_3}{b_3^2} \\
 -\frac{a_1}{b_1^2} & \frac{1}{b_1^2} & 0 & 0 \\
 -\frac{a_2}{b_2^2} & 0 & \frac{1}{b_2^2} & 0 \\
 -\frac{a_3}{b_3^2} & 0 & 0 & \frac{1}{b_3^2}
\end{array}
\right)\,,\nonumber\\
{\rm Re}\,\cal N&=\left(
\begin{array}{cccc}
 2 a_1 a_2 a_3 & -a_2 a_3 & -a_1 a_3 & -a_1 a_2 \\
 -a_2 a_3 & 0 & a_3 & a_2 \\
 -a_1 a_3 & a_3 & 0 & a_1 \\
 -a_1 a_2 & a_2 & a_1 & 0
\end{array}
\right)
\end{align}

This is the STU model, in the symplectic frame with prepotential
\be
F(X) = - \frac 16 C_{IJK}\frac{X^I X^J X^K}{X^0} = \frac{X^1 X^2 X^3}{X^0}\,,
\ee
Up to the sign of the superpotential, we identify the model \eqref{eq:4dAction} with $D_{IJK} =C_{IJK}$, provided we take the convention-dependent number
\begin{equation}
 c = \frac 1 {\sqrt{2}}
\end{equation}


\subsubsection{Convention matching of solutions}\label{ssec:matching}

The harmonic functions of 4D and 5D solutions are related as:
\be
H^0 = c V\,,\qquad H^I = c K^I\,,\qquad H_I = -c L_I\,,\qquad H_0 = 2 c M\,.
\ee
With $c = 1/\sqrt{2}$ in the harmonic functions. We apply same convention change to charges $\Gamma$ (note the conventional minus sign for the D2 charge).

After  applying this change of conventions, one sees that the probe potentials \eqref{eq:probe_denef}  and \eqref{eq:probe_iosif} exactly agree for supertubes. In 4D language, supertubes have the charge assignment of a D4-brane with world-volume flux. The only non-zero charges  are $p^3,q_1,q_2$ and  $q_0 \equiv q_1 q_2/d_3$.

\section{Particular Multi-center Solutions}\label{app:multicenter}
In this appendix, we give the explicit details of the backgrounds used in section \ref{ssec:mssusy} as well as the tunneling calculations in those backgrounds.

We write the solution in terms of explicit  harmonic functions with sources at $N$ centers $\vec r_i$:
\eal{ 
V &= \sum_{i = 1}^N \frac{v_{i}}{|\vec r - \vec r_i|}\,, \qquad \qquad &M &=m_0 + \sum_{i = 1}^N \frac{m_{0,i}}{|\vec r - \vec r_i|}\,,\\
K^I &= \sum_{i = 1}^N \frac{k^I_{i}}{|\vec r - \vec r_i|}\,, &L_I &= 1 + \sum_{i = 1}^N \frac{\ell_{I,i}}{|\vec r - \vec r_i|}\,.
}
We want to describe microstate geometries of the three-charge black hole in asymptotically flat five-dimensional spacetime. The only free parameters are the KK monopole charges $v_i$ and dipole charges $k^I_i$, as smoothness  at the different centers $\vec  r_i$ fixes the  sources of $L_I$ and $M$:
\be
\ell_{I,i} = -\tfrac 12 C_{IJK} \frac{k_i^J k_i^K}{v_i}\,,\qquad m_i = \frac 12 \frac{k_i^1 k_i^2 k_i^3}{q_i^2} \qquad \forall i \text{ (no sum)}\; .\label{eq:regularityharm}
\ee
Five-dimensional Minkowski asymptotics requires $\sum _{i=1}^N v_i = 1$ and fixes the constants of the harmonic functions by $V|_\infty  = K^I|_\infty = 0$, $L_I|_\infty  = 1$ and $M|_\infty = m_0$ with
\begin{equation}
 m_0 = \sum_{I=1}^3\sum_{i=1}^N k^I_i \,, \qquad \,.
\end{equation}

The physical, asymptotic charges are normalized as
\begin{equation}
Q_I \equiv \frac 1 {4\pi^2}\int Q_{IJ} \star _5 F^J\,,
\end{equation}
which gives asymptotically $Z_I = Q_I/\rho^2$ for the radius $\rho$ in standard polar coordinates on a constant time slice at infinity.

\subsection{Non-scaling solutions}\label{app:morenonscaling}

Here we  explain the background and calculations used in section \ref{sssec:nonscaling}.

\subsubsection{Background details}

Each GH center has three equal $k^I_i$ charges
\be k^I_i = -q_i N \hat{k} + \hat{k},\ee
so that $\sum_i k_i^I = 0$ and the physical flux between two centers is
\be \Pi_{ij}^I = (v_j-v_i) \hat{k}. \ee
The asymptotic charges in this background are given by:
\be Q\equiv Q_I = -4 \sum_j v_j (-v_j N \hat{k} + \hat{k})^2 = 4\hat{k}^2(N^2-1).\ee

The bubble equations simplify considerably for this simple system. The bubble equation for center $r_i$ gives us:
\be \label{eq:altsolbubble} 2\hat{k}^2 \sum_{j\neq i} \frac{v_j-v_i}{r_{ij}} = \frac32(-v_i N + 1),\ee
so that in particular, if $v_i=+1$:
\be 4\hat{k}^2 \sum_{j:v_j=-1}\frac{1}{r_{ij}}=\frac32(N-1),\ee
and if $v_i=-1$:
\be 4\hat{k}^2 \sum_{j:v_j=+1}\frac{1}{r_{ij}}=\frac32(N+1).\ee

We can also express the various harmonic functions in terms of $V = \sum_i v_i/r_i$ and $\tilde{V}=\sum_i 1/r_i$:
\begin{align}
 K^I &= (\tilde{V}-N V)\hat{k},\\
 L_I &= 1 - \left((N^2+1)V-2N\tilde{V}\right)\hat{k}^2,\\
 Z_I &= L + K^2/V = 1 + \frac{\hat{k}^2}{V}(\tilde{V}^2-V^2).
\end{align}
We note that:
\be Z_I(r_i) = \frac12(-1+3v_i N),\ee
where we used the bubble equations (\ref{eq:altsolbubble}).

We can see how $r_{12}$ scales with $N$, keeping $\hat{k}$ fixed. We do this by performing a linear fit on a log-log plot of $N$ vs. $r_{12}$ for the 6 data points of $N=11+8j$ for $j=1,\ldots,6$. The result is (inserting by dimensional-analysis the factor of  $k^2$):
\be r_{12}\sim \hat k^2 N^{-1.04}\,.\ee
The fit is very good, as can be seen in Figure \ref{fig:Nscalingr12} and by the statistical fit parameters\footnote{For the uninitiated, $R^2$ is the coefficient of determination and $p$ gives the so-called p-value.}: $1-R^2\approx (5\cdot 10^{-6})$ and $p\approx 10^{-11}$. However, if we increase $N$ much further, the exponent might still shift slightly to become $-1$.

\begin{figure}[h]
\begin{center}
 \includegraphics[width=0.5\textwidth]{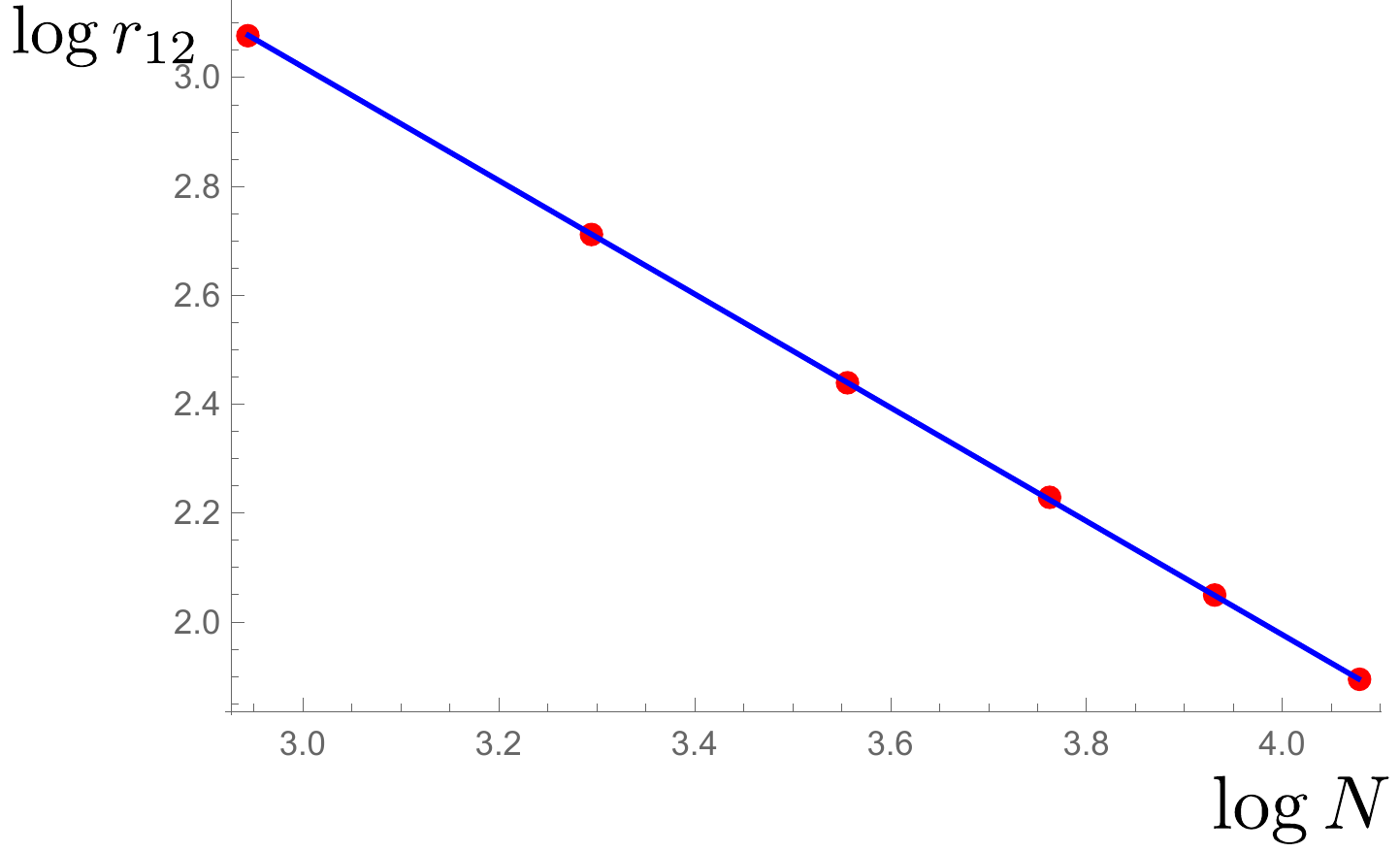}
 \end{center}
 \caption{A $\log-\log$ plot of $r_{12}$ versus $N$, keeping $\hat{k}$ fixed. The red dots are the actual data points and the blue line is the linear fit $r_{12}\sim N^{-1.04}$.}
 \label{fig:Nscalingr12}
\end{figure}

\subsubsection{Probe supertube}
We take $q_1^{\rm eff}(r_1)=0$, such that
\be q_1 = -d_3 \frac{k_1^2}{v_1} = d_3(N-1)\hat{k}.\ee
We also have $q_2^{\rm eff}(r_2)=0$, so that:
\be q_2 = -d_3 \frac{k_2^1}{v_2} = d_3(N+1)\hat{k}.\ee
Note that:
\begin{align}
q_1^{\rm eff}(r_2) &= -2d_3 \hat{k},\\
q_2^{\rm eff}(r_1) &= 2d_3 \hat{k}.
\end{align}

We take $q_2 = d_3(N+1)k$ so that the supersymmetric minimum between the first and second center is pushed all the way to the second center. We can wonder what happens when we take a different value for $q_2$, and thus move the supersymmetric minimum to the left: will the tunneling amplitude go up or down? We can take:
\be q_2 = d_3(N+1)\hat{k} + \lambda.\ee
An easy calculation shows that:
\be p(r) = | d_3 - \lambda \hat{k} (\tilde{V}-V)|.\ee
At $r_1$, we have (using the bubble equations):
\be \label{pri-difq2} p(r_1) = \left|1 - \frac34\frac{\lambda}{d_3 k} (N-1)\right|.\ee
Now, the expression that holds at the SUSY minimum $r_{SUSY}$ is:
\be \label{eq:STSUSYradius} d_3^2 \frac{Z_3}{V} = q_1^{\rm eff} q_2^{\rm eff}.\ee
This translates into an expression for $\lambda$ in function of the SUSY radius $r_{SUSY}$:
\be \lambda = \frac{d_3}{\hat{k}} (\tilde{V}-V)^{-1} = \frac{d_3}{2\hat{k}}\left(\sum_{j:v_j=-1}r_j^{-1}\right)^{-1},\ee
where of course $V,\tilde{V}$ are evaluated on $r=r_{SUSY}$. 
One can easily see that the function $\tilde{V}-V$ is everywhere positive, has local minima at the positions of all centers where $v_j=+1$ (as we are then furthest away from all the $v_j=-1$ centers), and goes to $+\infty$ at all centers where $v_j=-1$.  At the local minimum, the bubble equations give us:
\be (\tilde{V}-V)(r_j) = \frac3{4\hat{k}^2}(N-1),\ee
so that we have:
\be 0 \leq \frac{\lambda}{d_3 \hat k} < \frac43\frac{1}{N-1},\ee
for the allowed values of $\lambda$. The left boundary for $\lambda$ gives us $r_{SUSY}=r_2$ while approaching the right boundary in principle gives $r_{SUSY}\rightarrow r_1$. We see that clearly (\ref{pri-difq2}) satisfies:
\be p(r) \leq |d_3|.\ee
Moreover, we note that if $\lambda>0$ equation (\ref{eq:STSUSYradius}) implies that $p(r_{SUSY})=0$.

We can thus conclude that $p_{\lambda\neq0}(r_1)<p_{\lambda=0}(r_1)$ and moreover that $p_{\lambda\neq0}(r)$ is a strictly decreasing function between $r_1$ and $r_{SUSY}$ (reaching zero at $r_{SUSY}$) while $p_{\lambda=0}(r)$ is a constant function. Thus, the tunneling amplitude function $B$ we calculate with $\lambda=0$ will certainly give a strict \emph{upper bound} for $B$, namely $B_{\lambda\neq0}<B_{\lambda=0}$.

To calculate the tunneling amplitude for tunneling from $r_1$ to $r_2$, we first note that for this particular supertube probe:
\be p(r) = |d_3|,\ee
so that $p(r)$ is actually a constant function. Then, we simply have:
\be B = |d_3| r_{12},\ee
since we are tunneling between center $1$ and $2$.

\subsection{Scaling solutions}\label{app:morescaling}

We now  give more  details on the background of section \ref{sssec:scaling}.

\subsubsection{Background details}
The background has $N=4n+3$ centers with charges given by:
\begin{align}
 v_i &= (n\times\{20,-20\}, -12,25,-12,\{-20,20\}\times n),\\
 k^1_i &= \lambda^{-1}(n\times\{1375, -1325\}, \frac{5}{2} 12, \frac{5}{2}12, \frac52 12, \{-1325, 1375\}\times n),\\
 k^2_i &= \lambda^{-1}(n\times\{-\frac{1835}{2}+980 \hat{k}, \frac{1965}{2}-980\hat{k}\}, 12\hat{k}, 25\hat{k}, 12\hat{k},\\
 &\quad\{\frac{1965}{2}-980\hat{k}, -\frac{1835}{2}+980\hat{k}\}\times n),\\
 k^3_i &= \lambda^{-1}(n\times\{-\frac{8260}{3}, \frac{8380}{3}\}, \frac13 12, \frac13 25, \frac13 12, \{\frac{8380}{3}, -\frac{8260}{3}\}\times n),\\
\end{align}
and asymptotic constants
\be
(v_0, k_0^1,k_0^2,k_0^3, l_0^1,l_0^2,l_0^3,m_0) = (0,0,0,0,1,1,1,-\lambda^{-1}\frac12(\frac{833}{6}+49\hat{k}+310 n) ).
\ee
In this expressions, $\lambda$ is a parameter that will be determined by demanding that the charges remain (more or less) fixed when we vary $n$.

This solution always has a scaling solution for $\hat{k}\rightarrow k_*$ for a particular value of $k_*$. Since the bubble equations are:
\be \sum_{j\neq i}\frac{\langle \Gamma^i,\Gamma^j\rangle}{r_{ij}}=\langle h,\Gamma^i\rangle,\ee
a scaling solution is a solution to:
\be \sum_{j\neq i}\frac{\langle \Gamma^i,\Gamma^j\rangle}{r_{ij}}=0.\ee
We solved the latter equations to obtain the solution at the scaling point. This will always give us the exact value of $\hat{k}$ at the scaling point, but we are always free to rescale all of the $r_{ij}$ as $r_{ij}\rightarrow \epsilon\, r_{ij}$. In practice, we found the solutions up to $n=6$ ($N=23$).

We note that in these solutions (for $\lambda=1$), we have that $Q_I\sim N^2$ for large $N$. Numerically, the convergence to $N^2$ actually happens very slowly: for $N\sim100$ it is still about $\sim N^{1.96}$ for ($I=1,2$) or $\sim N^{2.02}$ (for $I=3$). For our practical purposes, this means that even by varying $\lambda$, we are not able to keep all three charges $Q_I$ exactly the same. Between the $n=1,2$ ($N=7,11$) solutions, the discrepancy between the charges is about $4-5\%$ for $Q_1$, $1-3\%$ for $Q_2$, $6-10\%$ for $Q_3$; the discrepancy does get smaller (the match is better) when $n$ increases. Still, the large $N$ behavior is $Q_I\sim N^2 k_i^2$, so that for example:
\be d_3 \sim q^{\rm eff}/k_i \sim N \frac{q^{\rm eff}}{Q^{1/2}},\ee
where $q^{\rm eff}$ is the physical effective charge of the supertube after the tunneling process.

\subsubsection{Rescaling distances}
We can use the scaling symmetry $r_{ij}\rightarrow \epsilon\, r_{ij}$ to rescale our solutions to have the same total microstate size. As a measure of the microstate size, we take the distance between the two outer centers. We normalize to the $n=1$ (7-center) solution.

One might worry whether keeping the total microstate size fixed is the right thing to do if one wants to compare different scaling microstates. Keeping the total charge  fixed, one could also compare the depth:
\be 
d=\int_{r_{\rm outer}}^{r_{\rm cut-off}} (V^3Z_1Z_2Z_3)^{1/6} dr,
\ee
of microstate solutions with different values of $n$. We find that the depths are approximately the same when we keep the total microstate size fixed (we get that the depth is at worst $93\%$ of the depth of the solution with $n=1$; moreover for $n$ large it is clear that this percentage gets much better -- comparing the solutions with $n=5$ and $n=6$ gives us 99,94\%).

As a side note, it appears  that $r_{12}$ (the distance between the two outer centers) converges to a constant value for large $n$, when we keep the charge and total microstate size fixed as described above. This may seem strange (since there are more and more centers, one would expect them to get more and more squashed), but what this is telling us is that when we increase $n$ and keep the total microstate size fixed, the inner centers get more and more squashed together while the outer centers remain approximately at the same distance.

\subsubsection{Tunneling amplitude}
Having fixed the charges and microstate size to be constant for different $N$, we can calculate $p(r)$ numerically in the usual way. Then, we can compare $\log N$ vs. $\log B$. We will ignore the $n=1$ ($N=7$) solution since $N$ is probably not quite large enough to be representative of the large-$N$ scaling. Fitting the last 5 datapoints ($n=2,\cdots,6$) gives us the scaling:
\be B \sim d_3 Q N^{-1.93}.\ee
The fit is very good, see Figure \ref{fig:scalingBwithN} (in the log-log plane): $1-R^2\approx 7.6\cdot 10^{-6}$ and $p\approx 9\cdot 10^{-9}$. We can try fitting only the last four datapoints, which also gives $N^{-1.93}$ behavior but with slightly ($4\cdot 10^{-6}$) lower $R^2$ and slightly ($1.5\cdot 10^{-6}$) worse $p$-value.

\begin{figure}[h]
\begin{center}
 \includegraphics[width=0.5\textwidth]{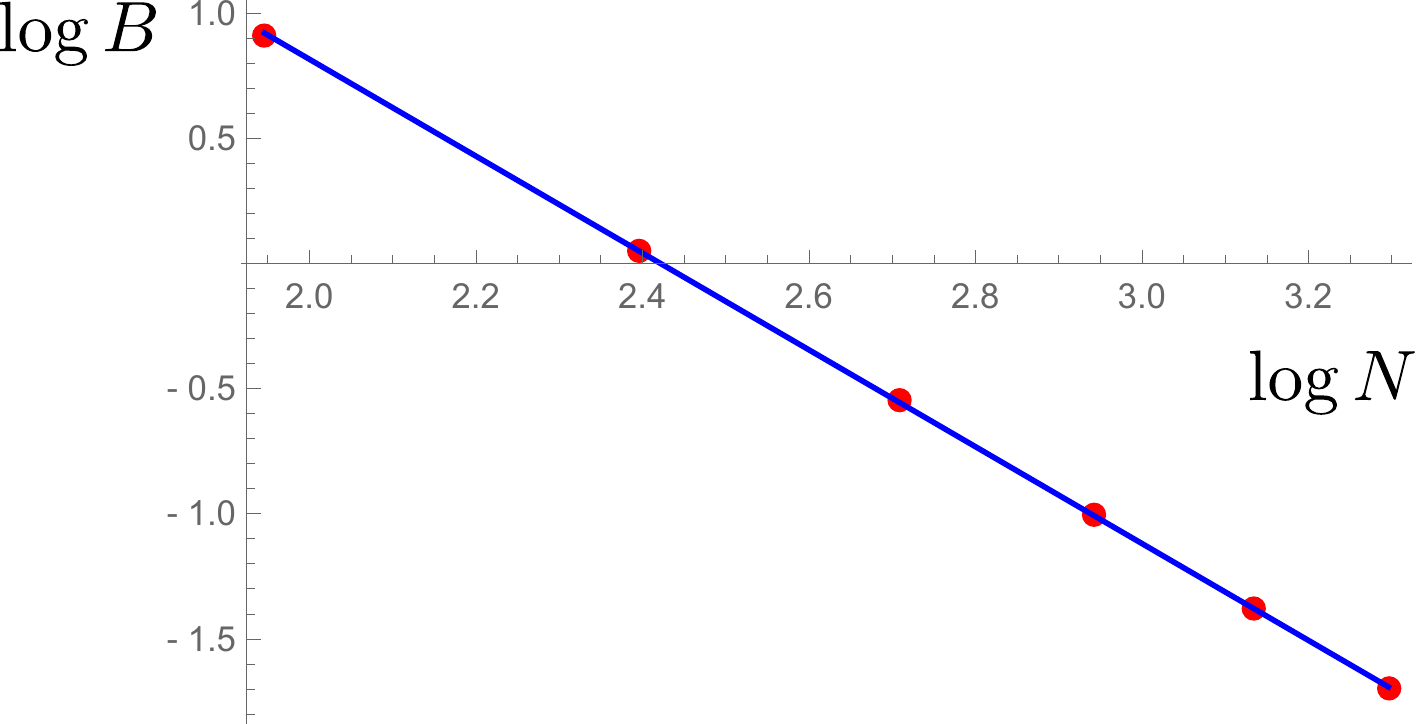}
 \end{center}
 \caption{A $\log-\log$ plot of $B$ versus $N$, keeping the charges $Q_I$ fixed. The red dots are the actual data points and the blue line is the linear fit $B\sim N^{-1.93}$.}
 \label{fig:scalingBwithN}
\end{figure}

Using $d_3\sim N q/Q^{1/2}$ then gives us:
\be B \sim q Q^{1/2} N^{-0.93}.\ee

\section{LLM Solutions}\label{app:LLM}
We summarize here the relevant formulas for computing tunneling rates in the LLM geometries. See~\cite{Lin:2004nb} and~\cite{Massai:2014wba} for more details.

\subsection{Background details}\label{app:LLMbackground}
The functions $h,G,A, \tilde{A}$ showing up in the M-theory and type IIA reduced LLM solutions of section~\ref{ssec:LLMbgd} are given by
\begin{equation}
h^{-2} = 2 y \cosh G \, , \quad G = {\rm arctanh}(2z)\,,
\quad A = \frac{z+\frac{1}{2}}{y^2}\, , \quad
\tilde{A}= \frac{z-\frac{1}{2}}{y^2}\, , 
\end{equation}
and $V$ is determined by the equations
\begin{equation}
y \partial_y V =\partial_x z\,, \quad y \partial_x V = - \partial_y z\, . \label{eq:V} 
\end{equation}
The full solution is determined in terms of a single master function $z(x,y)$ that obeys a linear equation:
\begin{equation}
 \partial_x^2 z + y \partial_y \left( \frac{\partial_y z}{y}
 \right) =0\, . \label{eq:z}
\end{equation}
%
%
%
A general smooth solution is determined by a superposition of solutions to \eqref{eq:V} and
\eqref{eq:z} with the boundary value of $z$ being either $1/2$ or  $-1/2$:
\begin{equation}
 z_0(x,y) = \frac{1}{2} \frac{x}{\sqrt{x^2+y^2}}\,,\quad
 V_0(x,y) = -\frac{1}{2} \frac{1}{\sqrt{x^2+y^2}}\,.
\end{equation}
For the metric \eqref{11metric} to asymptote to $AdS_4\times S^7$, the multi-strip solution must have a semi-infinite black region\footnote{In LLM jargon the regions where $z$ is $1/2$ are called ``white regions'' and those where it is $-1/2$ are called ``black regions.''} at one side of the $y=0$ line and a semi-infinite white region on the other. A general multi-strip solution is then obtained by superposition:
\begin{equation}
z(x,y)=\sum_{i=1}^{2s+1} (-1)^{i+1} z_0(x-x^{(i)},y)\,, \quad
V(x,y)= \sum_{i=1}^{2s+1} (-1)^{i+1} V_0(x-x^{(i)},y)\,,\label{multizV}
\end{equation}
where $x^{(i)}$ is the position of the $i$th boundary and $s$ denotes the number of pairs of white and black strips. For odd $i$ the boundary changes from black to white while for even $i$ the boundary changes from white to black. 
In the multi-strip solution \eqref{multizV} we get
\begin{eqnarray} 
c_3&=& \sum_{i=1}^{2s+1} (-1)^{i+1}
 \frac{2 (x-x^{(i)})^2+y^2}{2\sqrt{(x-x^{(i)})^2+y^2}}+x+ y^2 e^{2G} V +c\, ,\\
 c_5&=&\frac{2y^2}{1-2 z}- y^2+ c_3 H^{-1} h^{-2}V\, .
\end{eqnarray}
and similarly for $\tilde c_3$ and $\tilde c_5$.

\subsection{Probe strings and branes}\label{app:LLMprobeaction}

We give the details about the probe F1 strings and M2 branes used in section~\ref{ssec:LLMprobes}.

\subsubsection{Polarization potential}
The potential for an M5 brane with dissolved M2 charge can either be obtained directly in M-theory using the M5 brane potential of~\cite{Pasti:1997gx} or by dimensionally reducing the background and the probe to type IIA and exploring the action of a D4 brane wrapped on a three-sphere of the internal space and which carries $q$ units of dissolved F1 charge along $\omega_1$. The potential is given by~\cite{Massai:2014wba}
\begin{equation}\label{eq:LLMpotential}
V_{LLM}(x,y)= V_{DBI}(x,y)+V_{WZ}(x,y)\,,
\end{equation}
with the Dirac-Born-Infeld and Wess-Zumino actions
\begin{eqnarray}
 {V}_{DBI}(x,y)&=&=\mu_4 V_{S^3}H(x,y)^{-1} \sqrt{H(x,y) y^3 e^{3G(x,y)} + \left(\hat{q} - c_3(x,y)\right)^2}\,,\\
 {V}_{WZ}(x,y)&=&-\mu_4 V_{S^3}\left[\hat{q}{B}_{t\omega_1}(x,y) + c_5(x,y)\right]\,,
\end{eqnarray}
where for convenience we have defined a rescaled F1 string charge $\hat{q}=q/ 2\pi \alpha' \mu_4 V_{S^3}$ with $\mu_4=1/2\pi \alpha'(2\pi l_p)^3$ denoting the tension of the D4 brane and $V_{S^3}$ the volume of the wrapped three-sphere.
To avoid cumbersome notation we will also absorb factors of $\mu_4 V_{S^3}$ and denote the rescaled quantities with a hat.

One can show that the Hamiltonian has a minimum at $y=0$, where the master function $z$ takes the value $+1/2$. The potential~\eqref{eq:LLMpotential} becomes
\begin{equation}\label{hamplus}
\hat{V}_{LLM}^+ (x)=  H_+(x)^{-1} \sqrt{\frac{H_+(x)}{\zeta_+^3(x)} +
  \left(\hat{q}-c_3^+(x)\right)^2} - B_+(x)\left(\hat{q}-c_3^+(x)\right) -\frac{1}{\zeta_+^{2}(x)}  \, .
\end{equation}
The warp factor and the B field become
\begin{equation}\label{Hplus}
H_+(x) = \frac{\zeta_+^2(x) -V_+^2(x)}{\zeta_+(x)} \, ,\qquad B_+(x) =
-\frac{V_+(x)}{\zeta_+^2(x)-V_+^2(x)} \, ,
\end{equation}
and the functions entering in the RR potentials are
\begin{eqnarray}
c_3^+(x) &=&  \sum_{i=1}^{2s+1} (-1)^{i+1}|x-x^{(i)}|+ 
x + \frac{V_+(x)}{\zeta_+(x)^2} + c \, ,\label{eq:c3+}\\
c_5^+(x) &=& \frac{1}{\zeta_+^{2}(x)} - c_3^+(x) B_+(x)\,.
\end{eqnarray}
where the integration constant $c$ corresponds to a gauge choice.
In a multi strip solution we have
\begin{equation}
V_+(x) = -\frac{1}{2}\sum_{i=1}^{2s+1} \frac{(-1)^{i+1}}{|x-x^{(i)}|}\,, \qquad \zeta_+(x) = \frac12
\sqrt{\sum_{i=1}^{2s+1}(-1)^{i+1}\frac{|x-x^{(i)}|}{(x-x^{(i)})^3}}\, .
\end{equation}
The potential~\eqref{hamplus} has supersymmetric and metastable minima.

\paragraph{Supersymmetric minima.}
To satisfy $V_{LLM}^+=0$ we have to impose
\begin{equation}\label{susymin} 
 \left|c_3^+(x) - \frac{V_+(x)}{\zeta_+(x)^2}-\hat{q} \right| \zeta_+(x)-\left(c_3^+(x) - \frac{V_+(x)}{\zeta_+(x)^2}-\hat{q}\right)V_+(x)=0\, . 
 \end{equation}
There are two different ways to solve~\eqref{susymin} and, correspondingly, there exist two different kinds of minima: those where the probe M5 brane (D4 brane) shrinks to an M2 brane (F1 string), and those where it retains a finite-size. 
The first class of minima is obtained by noting that at the boundaries $x^{(i)}$ of the strips where both $S^3$ and $\tilde{S}^3$ shrink to zero size:
\begin{equation}
\lim_{x \to x^{(i)}} \frac{V_+(x)}{\zeta_+(x)}= (-1)^i \, .
\end{equation}
This means that the probe Hamiltonian can have degenerate supersymmetric minima located at the boundaries
$x^{(i)}$ if 
\begin{align}
\hat{q}^{\rm eff}_+(x^{(i)})&>0 \quad \textmd{($i$ odd)} \, \qquad \qquad {\rm or} \qquad \qquad \hat{q}^{\rm eff}_+(x^{(i)})<0  \quad
\textmd{(i even)}\, . \label{degeneratemin}
\end{align}
where we defined the effective M2 brane (F1 string) charge:
\begin{equation}\label{peff}
\hat{q}^{\rm eff}_+(x^{(i)})=\hat{q}-c_3^{+}(x^{(i)})\,.
\end{equation}

The second way to solve~\eqref{susymin} is to require the expression inside the absolute value and the brackets to vanish. 
The potential then has polarized supersymmetric minima inside a white strip $x^{(i)}<x_{susy}<x^{(i+1)}$ with $i$ odd, located at
\begin{eqnarray}\label{susy_minimum_exact}
x_{susy}&=& \frac{1}{2} \left( \hat{q}+ \left(\sum_{j=1}^i  - \sum_{j=i+1}^{2s+1}\right) (-1)^{j+1} x^{(j)} -c\right)\,,\nonumber\\
&=&\frac{1}{2} \left(\hat{q}+x^{(1)} +\Sigma_b^l - \Sigma_b^r-c\right)\, ,
\end{eqnarray}
where $\Sigma_b^l$ and $\Sigma_b^r$ are the total size of the black
strips that are respectively to the left and right of the white strip
in which the probe M5 brane polarizes. 
To have no Dirac strings at the left boundary of the white strip, $x^{(i)}$ with $i$ odd (``patch $i$''), we need 
\begin{equation}
0= c_3^+(x^{(i)})=c+x^{(i)}+\sum_{j=1}^{2s+1} (-1)^{j+1} |x^{(i)}-x^{(j)}|\,,
\end{equation}
 which implies
\begin{equation}\label{eq:patchi}
 c=\left(\sum_{j=1}^{i-1} -\sum_{j=i}^{2s+1} \right)(-1)^{j+1} x^{(j)}
 =\left(\sum_{j=1}^{i} -\sum_{j=i+1}^{2s+1} \right)(-1)^{j+1} x^{(j)}-2 x^{(i)}\,.
\end{equation}
In this gauge the location of the supersymmetric minimum takes the simple form
\begin{equation}
 x_{susy}=x^{(i)}+\frac{\hat{q}}{2}\,.
\end{equation}

\paragraph{Metastable minima.}
The potential also has metastable minima for $q<0$.\footnote{Note that for a different gauge choice $c$ there can be metastable minima for $q>0$.}
The full Hamiltonian~\eqref{hamplus} is well-approximated for small $x$ and small $|q|$ by:
\begin{equation}\label{Hpgeneral}
\hat{V}_{LLM}^+ \approx \hat{q}(a_1 +a_2x) - \frac{4}{\hat{q}}x^3 \, ,
\end{equation}
where
\begin{align}\label{coefficients}
  a_1 &=2\left(\sum_{j=1,j\neq i}^{2s+1}
   \frac{(-1)^{j}}{|x^{(i)}-x^{(j)}|}\right)^{-1}  \\
a_2 &= \frac{3}{4}
\left(\sum_{j=i+1}^{2s+1}\frac{(-1)^{j}}{(x^{(i)}-x^{(j)})^2}-\sum_{j=1}^{i-1}\frac{(-1)^{j}}{(x^{(i)}-x^{(j)})^2}\right)\left(a_1\right)^2
\, .
 \nonumber 
\end{align}
If $a_2>0$ the Hamiltonian~\eqref{Hpgeneral} always has a metastable minimum at
\begin{equation}\label{minima}
x=\frac{|\hat{q}|}{2} \sqrt{\frac{a_2}{3}} \, .
\end{equation}

\subsubsection{Tunneling amplitude}
Applying the derivation of the Euclidean momentum in section~\ref{sec:basics} to the IIA LLM background~\eqref{MIIAmetric}~-~\eqref{MIIAF4} we get:
\begin{equation}
 \hat{p}(x,y) =  \sqrt{(\hat{q}-c_3 (1+H^{-1} h^{-2} V^2)+ c_5 V)^2 - h^2 y^3 e^{3G} (1-y e^G)}\,,
\end{equation}
where we used~\eqref{eq:LLMpotential} and $g_{tt}=-H^{-1}$, $g_{xx}=h^2$.
In the $y\to0$ limit this becomes
\begin{equation}
 \hat{p}_+(x) =  |\hat{q}-c_3^+ + V_+/\zeta_+^2|\,.
\end{equation}
As a side note, equation~\eqref{susymin} can be written in terms of the momentum:
\begin{equation}
 |\hat{p}_+(x)| \zeta_+ -\hat{p}(x) V_+ = 0\,
\end{equation}
At supersymmetric minima we thus have
\begin{equation}\label{eq:pExsusy}
 \hat{p}_+(x_{susy})=0\,,
\end{equation}
while at a boundary we get
\begin{equation}
 \hat{p}_+(x^{(i)}) = |\hat{q}-c_3^+(x^{(i)})|\equiv|\hat{q}_{\rm eff}(x^{(i)})|\,.
\end{equation}
In a gauge where there are no Dirac strings at the boundary $x^{(i)}$ (given by $c_3^+(x^{(i)})=0$ or eq.~\eqref{eq:patchi}) we have that $\hat{q}_{\rm eff}(x^{(i)})=\hat{q}$ and hence $\hat{p}_+(x^{(i)})=|\hat{q}|$. When $\hat{q}=0$ we get degenerate supersymmetric minima located at boundaries and we recover~\eqref{eq:pExsusy}.

Using~\eqref{eq:c3+}, we can write the momentum more explicitly:
\begin{equation}\label{eq:pEcLLM}
 \hat{p}_+(x) = \left|\hat{q}- \sum_{i=1}^{2s+1} (-1)^{i+1} |x-x^{(i)}| - x -c \right|\,.
\end{equation}
To describe a BPS domain wall between the $x^{(i)}$ and $x^{(i+1)}$ or to compute the tunneling amplitude from a degenerate minimum at boundary $x^{(i)}$ to a polarized supersymmetric minimum inside the white strip $x^{(i)}<x_{susy}<x^{(i+1)}$
we can simplify~\eqref{eq:pEcLLM} to
\begin{equation}
 \hat{p}_+(x) =|\hat{q}-2 (x-x^{(i)})|\,,
\end{equation}
where we chose $c$ to take the value~\eqref{eq:patchi} so that there are no Dirac strings at the boundary $x^{(i)}$.
As a check, at the supersymmetric minimum $x_{susy}=x^{(i)}+\hat{q}/2$ we have $\hat{p}=0$.

\subsubsection{The energy of the metastable state}
As noted in section~\ref{ssec:LLMtunnel}, when the metastable state has only a small excess energy above the stable vacuum we can approximate the effective potential $V_{\rm eff}$ by the energy of the metastable F1 strings/M2 branes at the boundary $x^{(i)}$.
%
%
This can be obtained from the action
 \begin{equation}
  S_{F1} = -\frac{|q|}{2\pi \alpha'} \int dt d\omega_1 \sqrt{-g} +\frac{q}{2\pi \alpha'} \int dt d\omega_1  B_{t \omega_1}\,,
 \end{equation}
 where $\sqrt{-g}=H^{-1}$ and $B_{t \omega_1} =-H^{-1} h^{-2} V$. 
 Because the functions entering in the metric and the NS-NS B field are not the same, a probe F1 string (or M2 brane in the uplifted solution) is not BPS everywhere. 
The energy for one such \mbox{F1 string/anti-F1 string} ($|q|=1$) at the boundary $x^{(i)}$ is given by
\begin{equation}
 V_{F1}(x^{(i)})=  \frac{1}{2\pi \alpha'} \lim_{y\to0}\left[ H^{-1}\mp B_{t\omega_1}\right]\Big|_{x=x^{(i)}} = \frac{1}{2\pi \alpha'} [h^2\mp V]^{-1}\Big|_{x=x^{(i)}}\,.
 \label{eq:VF1}
\end{equation}
Uplifting this expression to M-theory gives the energy for one M2 brane/anti-M2 brane at the boundary $x^{(i)}$:
\begin{equation}
 V_{M2}(x^{(i)})= V_{F1}(x^{(i)})/2\pi R_{11}\,.
  \label{eq:VM2}
\end{equation}

The expression~\eqref{eq:VF1} vanishes for $q>0$ and $i$ odd and for $q<0$ and $i$ odd so that F1 strings are BPS at odd boundaries while anti-F1 strings are BPS at even boundaries, and similarly for the 11D uplift. Conversely, F1 strings and M2 branes at even boundaries while anti-F1 strings and anti-M2 branes at odd boundaries are metastable and their energy is given by, respectively, $V_{F1}(x^{(i)})$ and $ V_{M2}(x^{(i)})$. 

}
\providecommand{\href}[2]{#2}\begingroup\raggedright\endgroup

\end{document}